\documentclass[
 reprint,
 amsmath,amssymb,
 aps,
 prkeywords,
 nofootinbib
]{revtex4-2}

\usepackage[T1]{fontenc}
\usepackage{graphicx}
\usepackage{amsthm}
\usepackage{tikz}
\usepackage{xcolor}
\usetikzlibrary{arrows.meta,decorations.pathreplacing,positioning}
\makeatletter
\def\@hangfrom@appendix#1#2#3{%
 #1\@if@empty{#2}{#3}{#2\@if@empty{#3}{}{\quad #3}}%
}
\makeatother

\newtheorem{lemma}{Lemma}
\newtheorem{theorem}{Theorem}

\graphicspath{{./}}
\makeatletter
\AtBeginDocument{\let\selectlanguage\@gobble}
\makeatother

\begin{document}

\title{Energy-optimal scheduling for clocked reversible computing}

\author{Michael C. Thumann}
\email{thumann@utexas.edu}
\author{David Soloveichik}
\affiliation{The University of Texas at Austin, Austin, Texas 78712, USA}

\begin{abstract}
Computational dissipation may arise from both information erasure and finite-time operation, although both contributions can in principle be avoided in the absence of space and time constraints. Logically reversible computation can replace erasure with ancilla cleanup, and slow adiabatic protocols can eliminate any excess finite-time dissipation. Under limited memory and in finite time, however, choosing reversible cleanup over erasure requires additional dissipative operations, creating a semi-reversible energy--memory--time tradeoff. We formulate this tradeoff at a fixed operating point of an externally clocked thermally coupled computer, where reversible checkpoint moves and local irreversible erasures have specified effective prices $\epsilon_R$ and $\epsilon_I$. Their ratio $\lambda=\epsilon_I/\epsilon_R$ selects an energy-optimal semi-reversible schedule under a checkpoint-memory bound by trading reversible moves against erasures. For a sequential computation whose dependency graph is a line, we derive an exact separator recurrence and prove that an asymptotically optimal schedule repeats a strictly reversible block followed by one local erasure. Coupling this scheduling theorem to a stochastic-thermodynamic constitutive model shows how temperature, primitive duration, memory, and device-dependent operation costs determine whether the optimal schedule uses reversible cleanup, elective erasures, or erasure at every disposal.
\end{abstract}
\maketitle

\section{Introduction}

How little energy can a computation dissipate under physical
constraints? Koomey et al. found that computations per kilowatt-hour improved
by roughly a factor of two every 1.5 years for more than six decades
\cite{koomey_implications_2011}. That historical trend does not, however,
identify the constraints that will ultimately govern further
improvements. One such constraint is information erasure.

Landauer's foundational analysis showed that logically irreversible information
loss carries a minimum heat-dissipation cost at nonzero temperature
\cite{landauer_irreversibility_1961}. Although erasure may appear unavoidable
when repeated computations use a finite collection of gates and registers,
Bennett showed how uncomputing a computation after reversibly copying its
output restores the ancillas to their initial state, removing the need to
erase them \cite{bennett_logical_1973}.

Bennett's direct history-saving construction retains the intermediate states
needed to run the computation backward, so its memory use grows with the
computation length. Reversible pebbling reduces this requirement by
strategically interleaving computation, uncomputation, and recomputation. We consider line computations, which pass through a
sequence of complete states, each computed from just one predecessor state. Any Turing machine or logic circuit becomes a line computation when a
checkpoint (a saved copy of its complete state) is taken at fixed intervals of
steps or circuit layers \cite{bennett_timespace_1989}.
The line description ignores finer dependency structure, such as parts of the
state that can progress independently, which could be exploited by pebbling a directed acyclic graph (DAG), but we choose to analyze it for its generality. For a
line computation with $S$ working-state registers in addition to the
permanently available input, the farthest node admitting an erasure-free clean
finish, ending with only the output checkpoint present, is $2^{S-1}$
\cite{bennett_timespace_1989,knill_analysis_1995}. We call this limit the memory wall. At the wall, each additional pebble doubles the reachable length, while the number of reversible moves grows by nearly a factor of three\footnote{Bennett's construction uses three equal-cost recursive phases, giving an exact factor of three and leaving the extra pebble available in phase 1 unused. See Eq.~(\protect\ref{eq:knill_dp}).}, so avoiding erasure may quickly become expensive.

Thinking about reversible computing machines, Bennett distinguished ballistic, fully Brownian, and externally clocked Brownian regimes of reversible computation
\cite{bennett_thermodynamics_1982,bennett_notes_2003}. Ballistic machines
require isolation from thermal noise, and fully Brownian machines advance by a biased random walk with a
stochastic completion time. The externally clocked regime instead remains
coupled to a thermal environment while a control
protocol drives each operation over a fixed clock period. We focus on this latter regime and, to avoid confusion with
stochastic models, use the term ``externally clocked thermally coupled computer''. In
the slow adiabatic limit of an arbitrarily long clock period, logically
reversible operations approach zero dissipation
\cite{bennett_logical_1973,fredkin_conservative_1982}. Once a clock
period is specified, however, finite-time energetic costs must be supplied by
stochastic thermodynamics or another constitutive model
\cite{sivak_thermodynamic_2012,proesmans_finite-time_2020,
zhen_universal_2021,lee_speed_2022,kamijima_finite-time_2025}, and both
reversible steps and erasures incur excess work. Under bounded memory, avoiding
erasure requires additional reversible moves, so the energy-minimizing schedule must
balance reversible computation and uncomputation against erasure.

We formulate this line-graph scheduling problem for computations with uniform checkpoint width and uniform local operation prices. For a specified hardware operating point $H$, let $\epsilon_R(H)$ and
$\epsilon_I(H)$ be the effective prices of one reversible checkpoint move and
one local checkpoint erasure. For a schedule $\sigma$, let $R(\sigma)$ count
reversible moves and $I(\sigma)$ count local erasures. The fixed-price
compilation problem is
\begin{equation}
\min_{\sigma\in\mathcal F_{n,S}}
\left[
\epsilon_R(H)R(\sigma)+\epsilon_I(H)I(\sigma)
\right],
\label{eq:inner_optimization}
\end{equation}
where $\mathcal F_{n,S}$ is the feasible schedule set at line length $n$ and
checkpoint bound $S$. Once the operating point is fixed, the relative price
\[
\lambda_H=\frac{\epsilon_I(H)}{\epsilon_R(H)}
\]
sets the exchange rate between erasures and reversible moves, and it alone selects the optimal schedule.

A hardware-relaxed problem adds an outer minimization over the operating
point,
\begin{equation}
\min_H\min_{\sigma\in\mathcal F_{n,S}(H)}
\left[
\epsilon_R(H)R(\sigma)+\epsilon_I(H)I(\sigma)
\right].
\label{eq:outer_optimization}
\end{equation}
The feasible schedule set depends on $H$ only through the memory budget, which fixes
$S$. The outer minimization in Eq.~(\ref{eq:outer_optimization}) can lower the
total computational cost, but once the device is built, its protocols and
their durations are fixed, and Eq.~(\ref{eq:inner_optimization}) is the
remaining problem of choosing among valid schedules at given prices. For
example, in adiabatic superconducting logic, reversible
quantum-flux-parametron operations supply a finite-time reversible move cost \cite{takeuchi_reversibility_2017,yamae_minimum_2024},
while information-disposal operations supply a distinct
erasure cost
\cite{yamae_minimum_2024,herr_landauer-limited_2025}, and clocked quantum-dot cellular automata \cite{lent_device_1997,lent_bennett_2006} and fully adiabatic CMOS \cite{frank_reversible_2020} offer similar realizations. More generally, stochastic thermodynamics, device measurements, or another
physical model can supply the two effective prices without changing the
underlying scheduling problem.

The scheduling problem here builds on the work of Knill, who formulated the
time-optimal bounded-space reversible line game as a dynamic program
\cite{knill_analysis_1995}. Bennett's time--space tradeoff has also been
analyzed in Refs.~\cite{levine_note_1990,kralovic_time_2004}. Erasure as a
pebbling resource goes back to Li and Vit\'anyi, who traded time and space
against the number of irreversible erasures, taking erasures as the energy cost
of adiabatic computation \cite{li_reversibility_1996}, and to Li, Tromp, and
Vit\'anyi, who bounded how many pebbles a given number of erasures can save
\cite{li_reversible_1998}. Their springboard construction erases one checkpoint
per block of $2^{S-2}$ nodes and runs Bennett's recursive schedule within each
block. Measurement-based uncomputation plays a similar irreversible role in the
spooky pebble game \cite{kornerup_spooky_2025}. In these games, irreversible
operations are traded against space, and reversible moves count only toward run time. Here both carry prices at a fixed operating point, and we find the schedule that minimizes their weighted sum, with
checkpoint space bounded and erasure unconstrained. We call this priced pebble
game semi-reversible in the sense of Demaine et
al.~\cite{demaine_energy-efficient_2016}, who define the energy complexity of an
algorithm as the number of bits destroyed, counting reversible operations as
energetically free, as they are in the adiabatic limit. In that accounting,
erasure can save time or space but not energy, whereas in the priced pebble game, energy
optimization can favor erasure even when an erasure-free schedule fits within
the memory constraint.

We derive an exact separator recurrence for the priced pebble game and use it
to find the optimal schedule at any length. In a long computation, most
of the schedule lies far from the endpoints, in what we call the bulk. There the
recurrence reduces to choosing the best length for a strictly reversible block,
and we prove that an optimal schedule simply repeats that block, following each
with one erasure. Below $\lambda_H=1$, erasure at every disposal is optimal and additional checkpoint memory provides no energy
savings. As $\lambda_H$ increases, successive thresholds select longer reversible
blocks until the block reaches $2^{S-2}$ nodes, where it then coincides
with the springboard construction of Ref.~\cite{li_reversible_1998}, except with a
time-optimal schedule inside each block. Section~\ref{sec:physical_work} maps the
abstract prices to temperature, primitive duration, and device-dependent
finite-time coefficients using a stochastic-thermodynamic constitutive model.

\section{The priced pebble game as a semi-reversible computation}
\label{sec:priced_game}

We represent the saved states of a sequential computation by pebbles on the
line graph $\{0,1,\ldots,n\}$. Node $0$ is the permanently available input, and a pebble at node $i$ represents a stored checkpoint of the
corresponding computational state. At most $S$ pebbles may be active, excluding
node $0$.

The semi-reversible priced pebble game permits three moves
\cite{li_reversible_1998}:
\begin{itemize}
    \item A \emph{reversible computation} places a pebble on node $i$ at cost $\epsilon_R(H)$ only
    while node $i-1$ is pebbled.
    \item A \emph{reversible uncomputation} removes a pebble on node $i$ at cost $\epsilon_R(H)$ only
    while node $i-1$ is pebbled.
    \item An \emph{irreversible erasure} removes any pebble on node $i\neq n$ at cost $\epsilon_I(H)$,
    regardless of whether its predecessor is present.
\end{itemize}

A winning schedule starts with no active pebbles and ends with only node $n$
pebbled. In the fully reversible game, $S$ pebbles can reach node $2^S-1$ if intermediate pebbles may be left behind \cite{li_reversible_1998}, but a clean winning schedule is possible only up to the memory wall, $n\leq 2^{S-1}$.

Normalizing the prices in the game by $\epsilon_R(H)$ assigns unit cost to the
two reversible moves and the dimensionless price $\lambda_H$ to erasures.
Throughout, we assume that resets are local and that discarded checkpoints are uniformly distributed over their $2^b$ values, so that the quasistatic reset cost is $b k_B T\ln 2$. In the quasistatic limit, a fixed reset designed for uniformly distributed checkpoints costs $b k_B T\ln 2$ on any input, and its excess over the
generalized Landauer cost is dissipated as mismatch cost
\cite{kolchinsky_dependence_2021}. The generalized Landauer cost can be smaller. It depends on the checkpoint distribution, on correlations with other stored states \cite{wolpert_thermodynamics_2020,kolchinsky_dependence_2021}, and
on external side information \cite{faist_minimal_2015}. Once an algorithm and
input distribution are specified, algorithm- and data-specific resets may lower
the erasure price \cite{frank_foundations_2017} and shift the optimal schedule.

\begin{figure}
\centering
\resizebox{0.98\columnwidth}{!}{%
\begin{tikzpicture}[
    x=0.65cm,y=0.70cm,
    >=Stealth,
    wire/.style={line width=0.45pt, draw=black},
    livewire/.style={line width=0.68pt, draw=orange!75!black},
    lab/.style={font=\footnotesize},
    paneltitle/.style={font=\bfseries\footnotesize, align=center},
    gatelab/.style={font=\scriptsize},
    invgatelab/.style={font=\scriptsize,inner sep=0pt},
    erasegate/.style={draw, fill=black!8, line width=0.5pt}
]

%------------------------------------------
% Shared parameters
%------------------------------------------
\def\sep{0.060}
\def\gatehalf{0.45}
\def\gatepad{0.23}
\def\erasehalf{0.24}
\def\erasepad{0.19}
\def\zerotail{0.82}
\def\erasetail{0.72}
% label spacing
\def\rightlab{0.34}
\def\leftlab{0.34}
\def\inlabx{-0.22}

\def\yzero{0.0}
\def\yone{0.75}
\def\ytwo{1.50}
\def\ythree{2.25}
\def\yfour{3.00}

%------------------------------------------
% Shared helpers
%------------------------------------------
\newcommand{\pbundle}[4][wire]{%
  \foreach \k in {-1,0,1}{
    \draw[#1] ({#2},{#4+\k*\sep}) -- ({#3},{#4+\k*\sep});
  }%
}

\newcommand{\pgate}[4]{%
  \draw[draw=black, fill=white, line width=0.5pt]
    ({#1-\gatehalf},{#2-\gatepad}) rectangle ({#1+\gatehalf},{#3+\gatepad});
  \node[gatelab] at ({#1},{(#2+#3)/2}) {#4};
}
\newcommand{\pinvgate}[4]{%
  \draw[draw=black, fill=white, line width=0.5pt]
    ({#1-\gatehalf},{#2-\gatepad}) rectangle ({#1+\gatehalf},{#3+\gatepad});
  \node[invgatelab] at ({#1},{(#2+#3)/2}) {\resizebox{0.42cm}{!}{#4}};
}

\newcommand{\perase}[2]{%
  \draw[erasegate]
    ({#1-\erasehalf},{#2-\erasepad}) rectangle ({#1+\erasehalf},{#2+\erasepad});
  \node[gatelab] at ({#1},{#2}) {$E$};
}

%==========================================================
% Panel A: fully reversible cleanup
%==========================================================
\begin{scope}[shift={(0,10.35)}]
\node[paneltitle] at (.5,3.72) {(a) No erasure};

\def\xzero{1.00}
\def\xone{2.25}
\def\xtwo{3.50}
\def\xthree{4.75}
\def\xfour{6.00}
\def\xfive{7.25}
\def\xsix{8.50}
\def\xseven{9.75}
\def\xeight{11.00}
\def\xend{12.15}

\pbundle[livewire]{0.20}{\xend}{\yzero}
\node[lab] at (\xend+\rightlab,\yzero) {$\hat{x}$};
\node[lab] at (\inlabx,\yzero) {$\hat{x}$};

\pbundle{0.20}{\xzero-\gatehalf}{\yone}
\node[lab] at (\inlabx,\yone) {$0^b$};
\pbundle[livewire]{\xzero+\gatehalf}{\xone-\gatehalf}{\yone}
\pbundle[livewire]{\xone+\gatehalf}{\xtwo-\gatehalf}{\yone}
\pbundle{\xtwo+\gatehalf}{\xtwo+\zerotail}{\yone}
\node[lab] at (\xtwo+\zerotail+\rightlab,\yone) {$0^b$};

\pbundle{\xsix-\zerotail}{\xsix-\gatehalf}{\yone}
\node[lab] at (\xsix-\zerotail-\leftlab,\yone) {$0^b$};
\pbundle[livewire]{\xsix+\gatehalf}{\xseven-\gatehalf}{\yone}
\pbundle[livewire]{\xseven+\gatehalf}{\xeight-\gatehalf}{\yone}
\pbundle{\xeight+\gatehalf}{\xeight+\zerotail}{\yone}
\node[lab] at (\xeight+\zerotail+\rightlab,\yone) {$0^b$};

\pbundle{\xone-\zerotail}{\xone-\gatehalf}{\ytwo}
\node[lab] at (\xone-\zerotail-\leftlab,\ytwo) {$0^b$};
\pbundle[livewire]{\xone+\gatehalf}{\xthree-\gatehalf}{\ytwo}
\pbundle[livewire]{\xthree+\gatehalf}{\xfive-\gatehalf}{\ytwo}
\pbundle[livewire]{\xfive+\gatehalf}{\xseven-\gatehalf}{\ytwo}
\pbundle{\xseven+\gatehalf}{\xseven+\zerotail}{\ytwo}
\node[lab] at (\xseven+\zerotail+\rightlab,\ytwo) {$0^b$};

\pbundle{\xthree-\zerotail}{\xthree-\gatehalf}{\ythree}
\node[lab] at (\xthree-\zerotail-\leftlab,\ythree) {$0^b$};
\pbundle[livewire]{\xthree+\gatehalf}{\xfour-\gatehalf}{\ythree}
\pbundle[livewire]{\xfour+\gatehalf}{\xfive-\gatehalf}{\ythree}
\pbundle{\xfive+\gatehalf}{\xfive+\zerotail}{\ythree}
\node[lab] at (\xfive+\zerotail+\rightlab,\ythree) {$0^b$};

\pbundle{\xfour-\zerotail}{\xfour-\gatehalf}{\yfour}
\node[lab] at (\xfour-\zerotail-\leftlab,\yfour) {$0^b$};
\pbundle[livewire]{\xfour+\gatehalf}{\xend}{\yfour}
\node[lab] at (\xend+\rightlab,\yfour) {$\hat{y}$};

\pgate{\xzero}{\yzero}{\yone}{$U_1$}
\pgate{\xone}{\yone}{\ytwo}{$U_2$}
\pinvgate{\xtwo}{\yzero}{\yone}{$U_1^{-1}$}
\pgate{\xthree}{\ytwo}{\ythree}{$U_3$}
\pgate{\xfour}{\ythree}{\yfour}{$U_4$}
\pinvgate{\xfive}{\ytwo}{\ythree}{$U_3^{-1}$}
\pgate{\xsix}{\yzero}{\yone}{$U_1$}
\pinvgate{\xseven}{\yone}{\ytwo}{$U_2^{-1}$}
\pinvgate{\xeight}{\yzero}{\yone}{$U_1^{-1}$}
\end{scope}

%==========================================================
% Panel B: erase every checkpoint
%==========================================================
\begin{scope}[shift={(0,5.15)}]
\node[paneltitle] at (1.85,3.72) {(b) Erase every checkpoint};

\def\xuone{1.00}
\def\xutwo{2.65}
\def\xetwo{3.43}
\def\xuthree{4.30}
\def\xethree{5.08}
\def\xufour{5.95}
\def\xefour{6.73}
\def\xend{8.15}

\pbundle[livewire]{0.20}{\xend}{\yzero}
\node[lab] at (\xend+\rightlab,\yzero) {$\hat{x}$};
\node[lab] at (\inlabx,\yzero) {$\hat{x}$};

\pbundle{0.20}{\xuone-\gatehalf}{\yone}
\node[lab] at (\inlabx,\yone) {$0^b$};
\pbundle[livewire]{\xuone+\gatehalf}{\xutwo-\gatehalf}{\yone}
\pbundle[livewire]{\xutwo+\gatehalf}{\xetwo-\erasehalf-\erasetail}{\yone}
\pbundle[livewire]{\xetwo-\erasehalf-\erasetail}{\xetwo-\erasehalf}{\yone}
\pbundle{\xetwo+\erasehalf}{\xetwo+\zerotail}{\yone}
\node[lab] at (\xetwo+\zerotail+\rightlab,\yone) {$0^b$};

\pbundle{\xutwo-\zerotail}{\xutwo-\gatehalf}{\ytwo}
\node[lab] at (\xutwo-\zerotail-\leftlab,\ytwo) {$0^b$};
\pbundle[livewire]{\xutwo+\gatehalf}{\xuthree-\gatehalf}{\ytwo}
\pbundle[livewire]{\xuthree+\gatehalf}{\xethree-\erasehalf-\erasetail}{\ytwo}
\pbundle[livewire]{\xethree-\erasehalf-\erasetail}{\xethree-\erasehalf}{\ytwo}
\pbundle{\xethree+\erasehalf}{\xethree+\zerotail}{\ytwo}
\node[lab] at (\xethree+\zerotail+\rightlab,\ytwo) {$0^b$};

\pbundle{\xuthree-\zerotail}{\xuthree-\gatehalf}{\ythree}
\node[lab] at (\xuthree-\zerotail-\leftlab,\ythree) {$0^b$};
\pbundle[livewire]{\xuthree+\gatehalf}{\xufour-\gatehalf}{\ythree}
\pbundle[livewire]{\xufour+\gatehalf}{\xefour-\erasehalf-\erasetail}{\ythree}
\pbundle[livewire]{\xefour-\erasehalf-\erasetail}{\xefour-\erasehalf}{\ythree}
\pbundle{\xefour+\erasehalf}{\xefour+\zerotail}{\ythree}
\node[lab] at (\xefour+\zerotail+\rightlab,\ythree) {$0^b$};

\pbundle{\xufour-\zerotail}{\xufour-\gatehalf}{\yfour}
\node[lab] at (\xufour-\zerotail-\leftlab,\yfour) {$0^b$};
\pbundle[livewire]{\xufour+\gatehalf}{\xend}{\yfour}
\node[lab] at (\xend+\rightlab,\yfour) {$\hat{y}$};

\pgate{\xuone}{\yzero}{\yone}{$U_1$}
\pgate{\xutwo}{\yone}{\ytwo}{$U_2$}
\perase{\xetwo}{\yone}
\pgate{\xuthree}{\ytwo}{\ythree}{$U_3$}
\perase{\xethree}{\ytwo}
\pgate{\xufour}{\ythree}{\yfour}{$U_4$}
\perase{\xefour}{\ythree}
\end{scope}

%==========================================================
% Panel C: partially reversible
%==========================================================
\begin{scope}[shift={(0,0)}]
\node[paneltitle] at (1.1,3.72) {(c) Elective erasures};

\def\xzero{1.00}
\def\xone{2.25}
\def\xtwo{3.50}
\def\xthree{4.75}
\def\xfour{6.00}
\def\xfive{7.25}
\def\xerase{8.25}
\def\xend{9.45}

\pbundle[livewire]{0.20}{\xend}{\yzero}
\node[lab] at (\xend+\rightlab,\yzero) {$\hat{x}$};
\node[lab] at (\inlabx,\yzero) {$\hat{x}$};

\pbundle{0.20}{\xzero-\gatehalf}{\yone}
\node[lab] at (\inlabx,\yone) {$0^b$};
\pbundle[livewire]{\xzero+\gatehalf}{\xone-\gatehalf}{\yone}
\pbundle[livewire]{\xone+\gatehalf}{\xtwo-\gatehalf}{\yone}
\pbundle{\xtwo+\gatehalf}{\xtwo+\zerotail}{\yone}
\node[lab] at (\xtwo+\zerotail+\rightlab,\yone) {$0^b$};

\pbundle{\xone-\zerotail}{\xone-\gatehalf}{\ytwo}
\node[lab] at (\xone-\zerotail-\leftlab,\ytwo) {$0^b$};
\pbundle[livewire]{\xone+\gatehalf}{\xthree-\gatehalf}{\ytwo}
\pbundle[livewire]{\xthree+\gatehalf}{\xfive-\gatehalf}{\ytwo}
\pbundle[livewire]{\xfive+\gatehalf}{\xerase-\erasehalf-\erasetail}{\ytwo}
\pbundle[livewire]{\xerase-\erasehalf-\erasetail}{\xerase-\erasehalf}{\ytwo}
\pbundle{\xerase+\erasehalf}{\xerase+\zerotail}{\ytwo}
\node[lab] at (\xerase+\zerotail+\rightlab,\ytwo) {$0^b$};

\pbundle{\xthree-\zerotail}{\xthree-\gatehalf}{\ythree}
\node[lab] at (\xthree-\zerotail-\leftlab,\ythree) {$0^b$};
\pbundle[livewire]{\xthree+\gatehalf}{\xfour-\gatehalf}{\ythree}
\pbundle[livewire]{\xfour+\gatehalf}{\xfive-\gatehalf}{\ythree}
\pbundle{\xfive+\gatehalf}{\xfive+\zerotail}{\ythree}
\node[lab] at (\xfive+\zerotail+\rightlab,\ythree) {$0^b$};

\pbundle{\xfour-\zerotail}{\xfour-\gatehalf}{\yfour}
\node[lab] at (\xfour-\zerotail-\leftlab,\yfour) {$0^b$};
\pbundle[livewire]{\xfour+\gatehalf}{\xend}{\yfour}
\node[lab] at (\xend+\rightlab,\yfour) {$\hat{y}$};

\pgate{\xzero}{\yzero}{\yone}{$U_1$}
\pgate{\xone}{\yone}{\ytwo}{$U_2$}
\pinvgate{\xtwo}{\yzero}{\yone}{$U_1^{-1}$}
\pgate{\xthree}{\ytwo}{\ythree}{$U_3$}
\pgate{\xfour}{\ythree}{\yfour}{$U_4$}
\pinvgate{\xfive}{\ytwo}{\ythree}{$U_3^{-1}$}
\perase{\xerase}{\ytwo}
\end{scope}

\end{tikzpicture}
}%
\caption{\label{pebbles}
Three $n=4$ schedules. Each wire carries one node, orange wires are live checkpoint bits, and black bundles marked $0^b$ are blank ancillas. Panel (a) uses reversible uncomputation, panel (b) erases each old checkpoint immediately, and panel (c) uses one elective erasure.}
\end{figure}
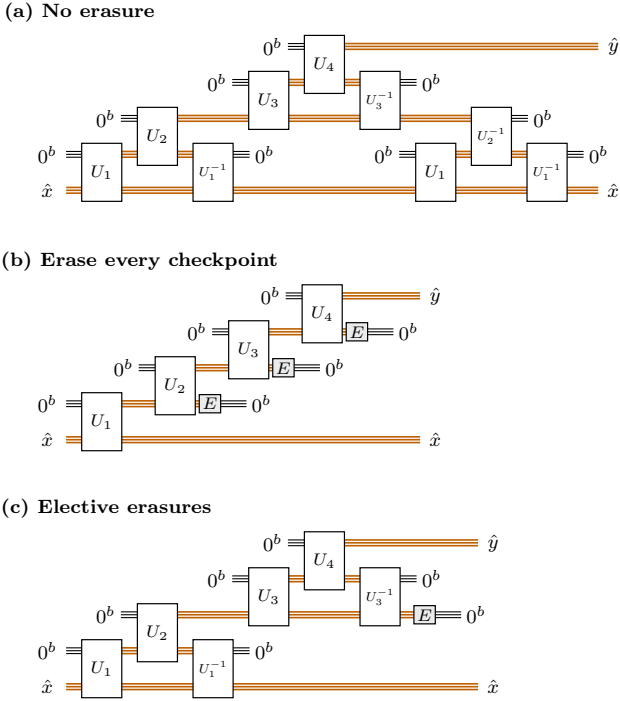

From here on, we fix the hardware operating point and write $\lambda=\lambda_H$. To connect the pebble game to individual bits and reversible, conservative circuits \cite{fredkin_conservative_1982}, we assume that every checkpoint
contains the same number of bits, $b$. A working-memory budget of $M$ bits then
allows \[S=\left\lfloor\frac{M}{b}\right\rfloor\] simultaneously active checkpoints. Figure~\ref{pebbles} illustrates the
circuit-level interpretation of fully reversible, semi-reversible, and fully
irreversible schedules. The mixed schedule in panel (c) shows that well placed erasures can reduce the number of reversible steps.

\section{Exact schedule optimization at fixed prices}

With the game and its prices fixed, let $C_\lambda(n,S)$ be the minimum dimensionless cost of a winning priced schedule on an
interval of length $n$ with space bound $S$. The base cases are
\begin{equation}
C_\lambda(1,S)=1,\qquad C_\lambda(n,1)=\infty\quad(n\ge2).
\label{eq:C_base}
\end{equation}
The first case is the forced placement of node $1$, while the second expresses the fact that
advancing beyond node $1$ requires simultaneously holding a predecessor and a new pebble.

Knill's dynamic program for the bounded-space fully reversible pebble game \cite{knill_analysis_1995} gives the recurrence
\begin{multline}
F(n,S) = \min_{1\le m<n} \Big[
F(m,S)+F(n-m,S-1)+\\
F(m,S-1) \Big].
\label{eq:knill_dp}
\end{multline}
where $F(n,S)$ is the minimum number of reversible moves in the strictly reversible schedule. The base cases are again $F(1,S)=1$ and $F(n,1)=\infty$ for $n\ge2$. The three terms in Eq.~(\ref{eq:knill_dp}) correspond to the three-phase schedule \[\emptyset \xrightarrow[\text{Phase 1}]{} \{m\} \xrightarrow[\text{Phase 2}]{} \{m,n\} \xrightarrow[\text{Phase 3}]{} \{n\}\] in which an intermediate node $m$ is chosen as the furthest intermediate node that the lowest pebble achieves at any time. An optimal schedule need not run these phases one after another, since moves below and above $m$ can interleave. Knill showed, however, that projecting the moves onto the nodes below and above $m$ splits any optimal schedule into the three subproblems, so the phases can always be run in sequence at no extra cost \cite{knill_analysis_1995}. In sequential form, phase 1 uses only nodes below $m$, phase 2 only nodes above $m$, and phase 3 again only nodes below $m$. Phases 1 and 2 are recursive subproblems with $m$ as the output and the input, respectively, and phase 3 is the time reverse of a schedule that builds $m$. Phase 1 has all $S$ pebbles available. Phases 2 and 3 have only $S-1$, because the pebble on $m$ is held as the input during phase 2 and the pebble on $n$ is held as the output during phase 3.

We prove that the priced pebble game inherits this separator structure, but phase 3 can now remove $m$ either by reversible uncomputation at cost $F(m,S-1)$ or by erasure at cost $\lambda$. Theorem~\ref{thm:priced_separator_recurrence} takes the cheaper of the two. Erasures inside phases 1 and 2 appear as phase-3 choices at deeper levels of the recurrence.

\begin{theorem}[Priced separator recurrence]
\label{thm:priced_separator_recurrence}
For \(n\ge 2\) and \(S\ge 2\),
\begin{multline}
C_\lambda(n,S) = \min_{1\le m<n} \left[ C_\lambda(m,S) + C_\lambda(n-m,S-1) \right. \\
\left. + \min\{F(m,S-1),\lambda\} \right].
\label{eq:priced_dp}
\end{multline}
\end{theorem}

The lower bound requires the projection argument to survive erasure. Appendix~\ref{app:normal_form} proves that any optimal priced schedule can be transformed into a valid projection form, and thus a sequential three-phase form without increasing its cost or move count (Lemmas~\ref{lem:clean_separator_normal_form} and~\ref{lem:separator_normal_form}). 

\begin{proof}
For the upper bound, fix \(m\in\{1,\ldots,n-1\}\). In phase 1, compute node \(m\) on
the lower interval \(\{1,\ldots,m\}\) using an optimal priced schedule with
space \(S\). This costs $C_\lambda(m,S)$. In phase 2, hold \(m\) fixed and compute the upper interval
\(\{m+1,\ldots,n\}\), treating \(m\) as the permanent predecessor of
\(m+1\). After reindexing, this is a priced schedule of length \(n-m\) using at
most \(S-1\) active pebbles, and so costs $C_\lambda(n-m,S-1)$.

At this point the only active pebbles are \(m\) and \(n\). In phase 3, dispose of \(m\).
Either erase \(m\) directly at cost \(\lambda\), or reversibly uncompute the
lower block with the remaining pebbles while \(n\) is held fixed. Run backwards, reversible cleanup is a schedule that builds \(m\) with \(S-1\) pebbles, so the cheapest reversible cleanup costs $F(m,S-1)$. It follows that for every \(m\),
\begin{multline*}
C_\lambda(n,S)
\le
C_\lambda(m,S)
+C_\lambda(n-m,S-1)\\
+\min\{F(m,S-1),\lambda\}.
\end{multline*}
Minimizing over \(m\) gives the upper bound. The lower bound is the statement of
Lemma~\ref{lem:separator_normal_form}, proven in Appendix~\ref{app:normal_form}.

Combining the upper and lower bounds proves the recurrence.
\end{proof}

For fixed $\lambda$, tabulating the recurrence through line length $N$ and
checkpoint budget $S$ requires $O(N^2S)$ time and $O(NS)$ storage, including
the table for $F$. When several schedules have the same energy, we report
one with the smaller move count \(R+I\). If that quantity also ties, we
report the schedule with more erasures. Neither convention
changes the minimum cost.

\begin{figure*}[t]
\centering
\includegraphics[width=0.49\textwidth]{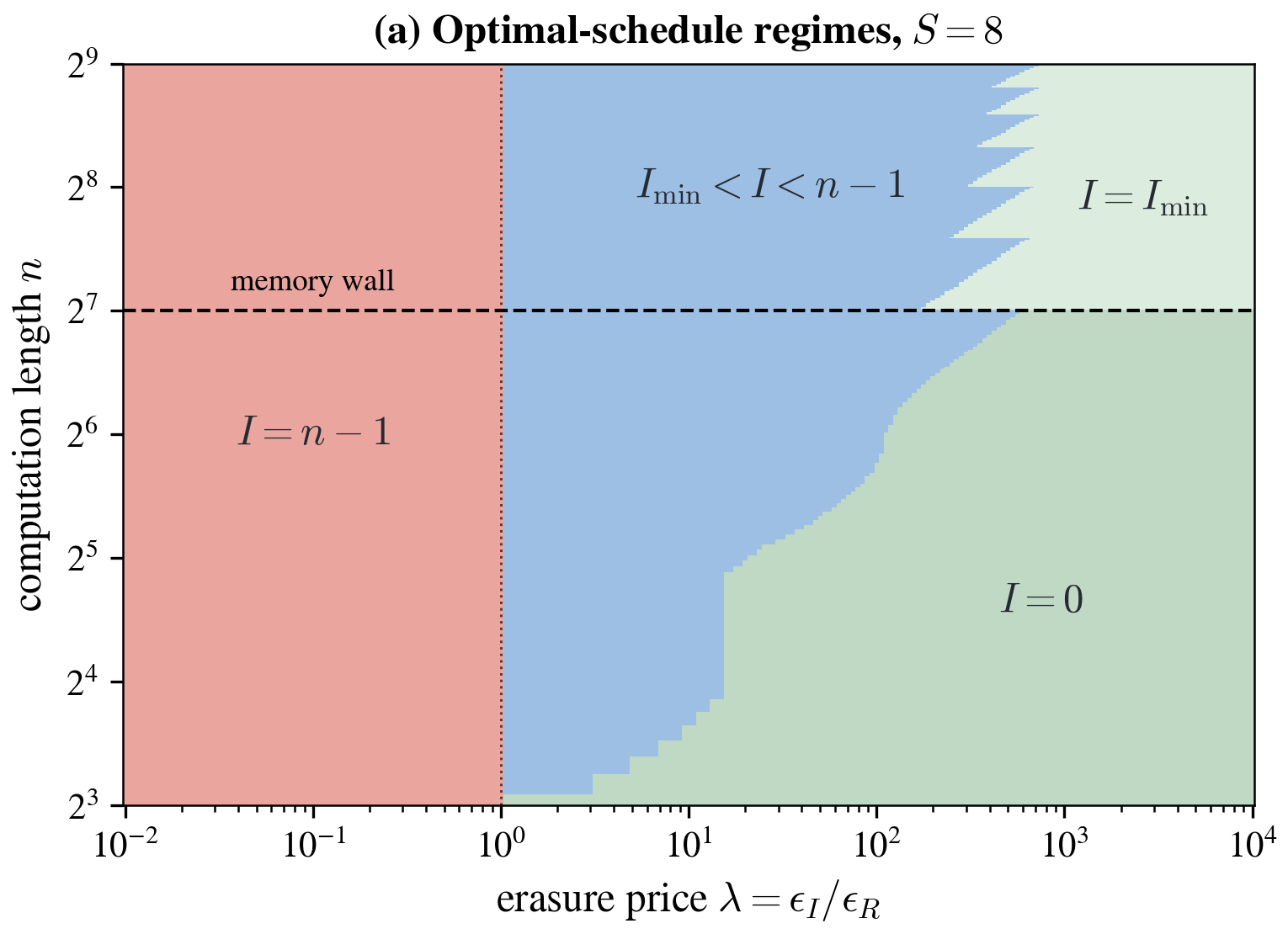}\hspace{0.015\textwidth}%
\includegraphics[width=0.49\textwidth]{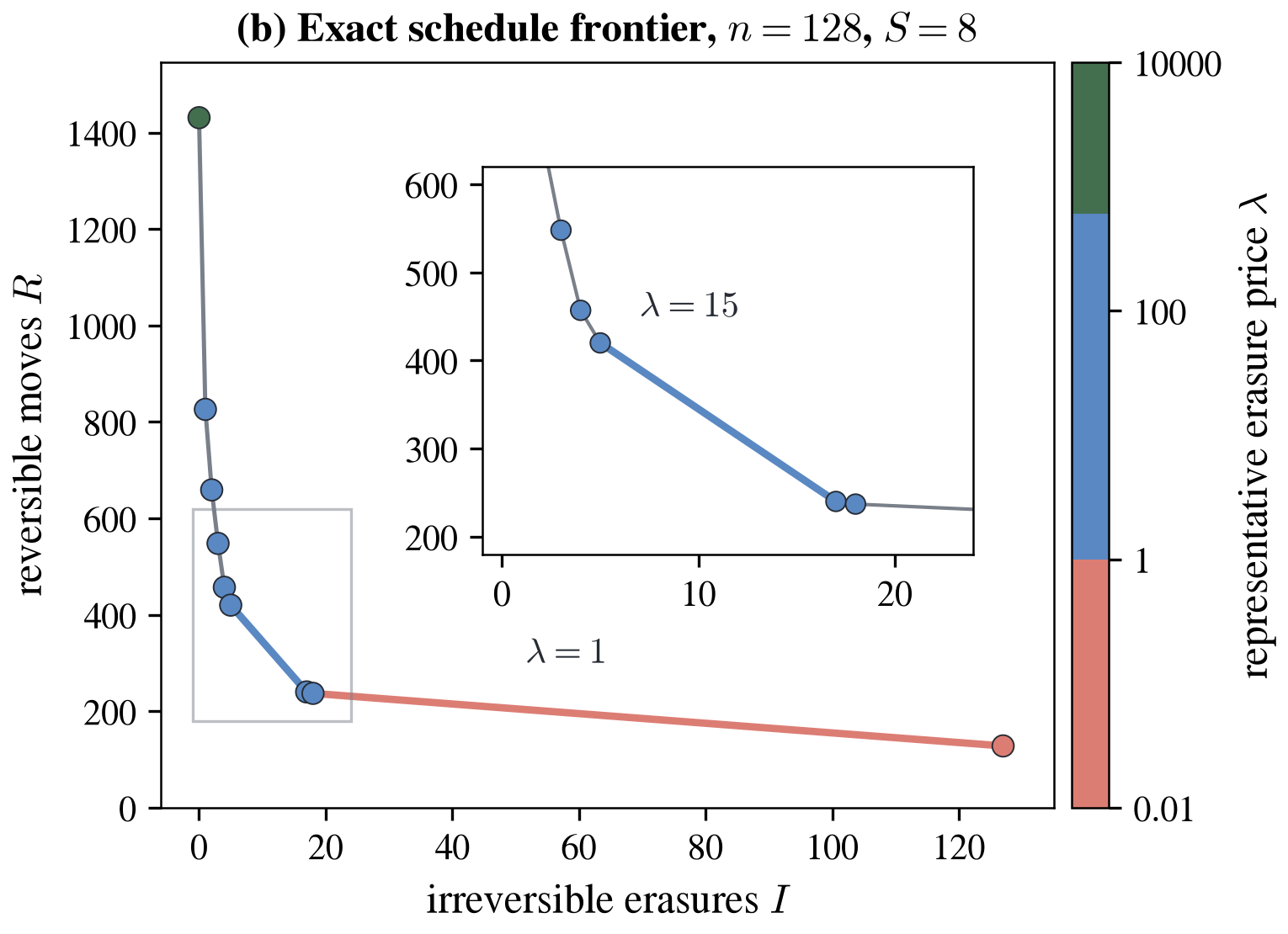}

\caption{\label{fig:regimes_pareto}%
Exact schedule structure at $S=8$. (a)  Optimal erasure count $I$ in the $(n,\lambda)$ plane, from the exact
recurrence on a 241-point logarithmic price grid, with $I_{\min}$ as defined
in the text. The dashed line marks the memory wall $n=2^{S-1}=128$. (b) Supported reversible-move/erasure frontier at the memory wall $n=128$. Color indicates a representative price $\lambda$ that selects each supported $(I,R)$ schedule.}
\end{figure*}

\section{Finite-length schedule regimes}
\label{sec:limiting}

The recurrence Eq.~(\ref{eq:priced_dp}) recovers two limiting regimes in
closed form. First consider the case
\(S=2\). For \(n\ge 3\), the only feasible separator is
\[
m=n-1,
\]
giving the optimal schedule as a two-pebble erasure glider. The costs are
\[
C_\lambda(1,2)=1,
\qquad
C_\lambda(2,2)=2+\min\{1,\lambda\}.
\]
For \(n\ge 3\), the recurrence leads to 
\[
C_\lambda (n,2) = \begin{cases} 
      n+1+(n-2)\lambda & \text{if } \lambda\ge 1 \\
       n+(n-1)\lambda  & \text{if }\lambda\le 1.
\end{cases}
\]
Since any two-pebble schedule remains legal when more
pebbles are available, \(C_\lambda(n,S)\le C_\lambda(n,2)<\infty\) for all
\(n\) and \(S\ge2\). Unlike the strictly reversible game, the priced game
faces no memory wall.

At the opposite extreme, suppose \(S\ge n\) and \(\lambda\ge 1\). Then the
strictly reversible full compute-uncompute schedule is feasible at the optimal cost $2n-1$. In the separator recurrence this regime is the minimizing
choice \(m=1\) at every level.

Between these extremes, limited memory forces a choice between recomputation and erasure. Whenever $S<n$ pebbles are available, and the output lies beyond the memory wall, $n>2^{S-1}$, every winning schedule contains at least one erasure. Let $I_{\min}(n,S)$ be the fewest
erasures of any winning schedule, so that $I_{\min}=0$ up to the wall and
$I_{\min}\ge1$ beyond it. Price and memory then divide the $(n,\lambda)$ plane into the regimes of Fig.~\ref{fig:regimes_pareto}(a), labeled by the optimal erasure count $I$. For $\lambda<1$, erasure is cheaper than reversible removal, so every disposal is erased and the optimum is the two-pebble erasure glider with $I=n-1$. For $\lambda>1$, any erasures beyond $I_{\min}$ are elective. They are chosen on energetic grounds even though a schedule with fewer erasures exists. The optimal erasure count never increases with $\lambda$, and for large enough $\lambda$ it reaches $I=I_{\min}$. There the optimum is the minimum-erasure schedule, which is erasure-free below the wall and contains only the number of unavoidable erasures above it.

The same tradeoff can be viewed in move-count space. Let \(\mathcal A_{n,S}\) be the set of attainable pairs \((I(\sigma),R(\sigma))\). Figure~\ref{fig:regimes_pareto}(b) shows its supported frontier. As \(\lambda\) increases, the selected point moves toward fewer erasures and more reversible work.

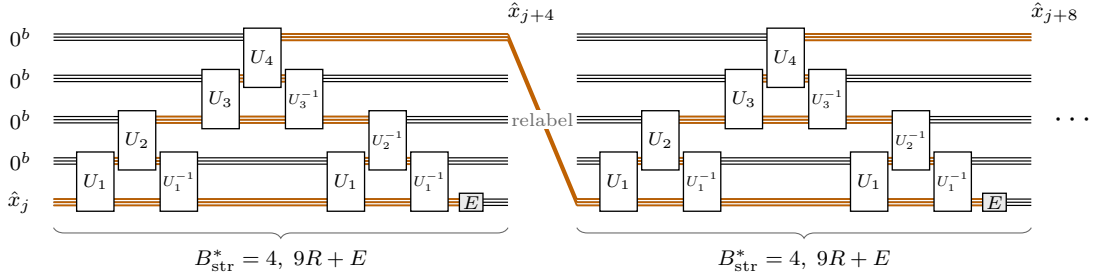
\begin{figure*}[t]
\centering
\begin{tikzpicture}[
    x=0.65cm,y=0.70cm,
    >=Stealth,
    wire/.style={line width=0.45pt,draw=black},
    livewire/.style={line width=0.68pt,draw=orange!75!black},
    lab/.style={font=\footnotesize},
    gatelab/.style={font=\scriptsize},
    invgatelab/.style={font=\scriptsize,inner sep=0pt},
    note/.style={font=\scriptsize,color=black!58},
    erasegate/.style={draw,fill=black!8,line width=0.5pt}
]
\def\bsep{0.060}
\def\rowstep{0.78}
\def\dx{0.85}
\def\gatehalf{0.38}
\def\gatepad{0.17}
\def\erasehalf{0.24}
\def\erasepad{0.17}

\newcommand{\bbundle}[4][wire]{%
  \foreach \k in {-1,0,1}{%
    \draw[#1] ({#2},{#4+\k*\bsep}) -- ({#3},{#4+\k*\bsep});
  }%
}
\newcommand{\bgate}[4]{%
  \draw[draw=black,fill=white,line width=0.5pt]
    ({#1-\gatehalf},{#2-\gatepad})
    rectangle
    ({#1+\gatehalf},{#3+\gatepad});
  \node[gatelab] at ({#1},{(#2+#3)/2}) {#4};
}
\newcommand{\binvgate}[4]{%
  \draw[draw=black,fill=white,line width=0.5pt]
    ({#1-\gatehalf},{#2-\gatepad})
    rectangle
    ({#1+\gatehalf},{#3+\gatepad});
  \node[invgatelab] at ({#1},{(#2+#3)/2}) {\resizebox{0.42cm}{!}{#4}};
}
\newcommand{\berase}[2]{%
  \draw[erasegate]
    ({#1-\erasehalf},{#2-\erasepad})
    rectangle
    ({#1+\erasehalf},{#2+\erasepad});
  \node[gatelab] at ({#1},{#2}) {\(E\)};
}
\newcommand{\strictfour}[2]{%
  \foreach \j in {0,...,4}{%
    \bbundle{#1}{#2}{\j*\rowstep}
  }
  \bbundle[livewire]{#1}{#1+10*\dx}{0}
  \bbundle[livewire]{#1+1*\dx}{#1+3*\dx}{1*\rowstep}
  \bbundle[livewire]{#1+7*\dx}{#1+9*\dx}{1*\rowstep}
  \bbundle[livewire]{#1+2*\dx}{#1+8*\dx}{2*\rowstep}
  \bbundle[livewire]{#1+4*\dx}{#1+6*\dx}{3*\rowstep}
  \bbundle[livewire]{#1+5*\dx}{#2}{4*\rowstep}
  \bgate{#1+1*\dx}{0}{1*\rowstep}{\(U_1\)}
  \bgate{#1+2*\dx}{1*\rowstep}{2*\rowstep}{\(U_2\)}
  \binvgate{#1+3*\dx}{0}{1*\rowstep}{\(U_1^{-1}\)}
  \bgate{#1+4*\dx}{2*\rowstep}{3*\rowstep}{\(U_3\)}
  \bgate{#1+5*\dx}{3*\rowstep}{4*\rowstep}{\(U_4\)}
  \binvgate{#1+6*\dx}{2*\rowstep}{3*\rowstep}{\(U_3^{-1}\)}
  \bgate{#1+7*\dx}{0}{1*\rowstep}{\(U_1\)}
  \binvgate{#1+8*\dx}{1*\rowstep}{2*\rowstep}{\(U_2^{-1}\)}
  \binvgate{#1+9*\dx}{0}{1*\rowstep}{\(U_1^{-1}\)}
  \berase{#1+10*\dx}{0}
}
\newcommand{\periodrelabel}[2]{%
  \foreach \k in {-1,0,1}{%
    \draw[livewire]
      ({#1},{4*\rowstep+\k*\bsep}) -- ({#2},{\k*\bsep});
  }
  \node[note,fill=white,inner sep=1pt]
    at ({(#1+#2)/2},{2*\rowstep}) {relabel};
}

\def\xA{1.20}
\def\xB{10.45}
\def\xC{11.85}
\def\xD{21.10}

\node[lab,anchor=east] at (\xA-0.28,0) {\(\hat{x}_j\)};
\foreach \j in {1,...,4}{%
  \node[lab,anchor=east] at (\xA-0.28,\j*\rowstep) {\(0^b\)};
}

\strictfour{\xA}{\xB}
\node[lab,anchor=south west] at (\xB-0.18,4*\rowstep+0.10)
  {\(\hat{x}_{j+4}\)};
\periodrelabel{\xB}{\xC}
\strictfour{\xC}{\xD}
\node[lab,anchor=south west] at (\xD-0.18,4*\rowstep+0.10)
  {\(\hat{x}_{j+8}\)};
\node[font=\large,anchor=west] at (\xD+0.28,1.56) {\(\cdots\)};

\draw[decorate,decoration={brace,mirror,amplitude=4pt},draw=black!60]
  (\xA,-0.48) -- (\xB,-0.48)
  node[midway,below=5pt,lab]
  {\(B_{\rm str}^*=4,\;9R+E\)};
\draw[decorate,decoration={brace,mirror,amplitude=4pt},draw=black!60]
  (\xC,-0.48) -- (\xD,-0.48)
  node[midway,below=5pt,lab]
  {\(B_{\rm str}^*=4,\;9R+E\)};
\end{tikzpicture}
\caption{\label{fig:bulk_schedule}
Two interior periods of the optimal bulk schedule for $S=4$ and $\lambda=8$,
where $B_{\rm str}^*=4$ and $F(4,3)=9$. Each period uses at most four live
pebbles, performs nine reversible moves, erases the old boundary once, and
relabels the output as the next input. Thus $q$ interior periods have $R=9q$,
$I=q$, and normalized cost $17q$. The protected global input and final output
contribute only a bounded endpoint correction.}
\end{figure*}

\begin{figure*}
    \centering
    \includegraphics[width=\linewidth]{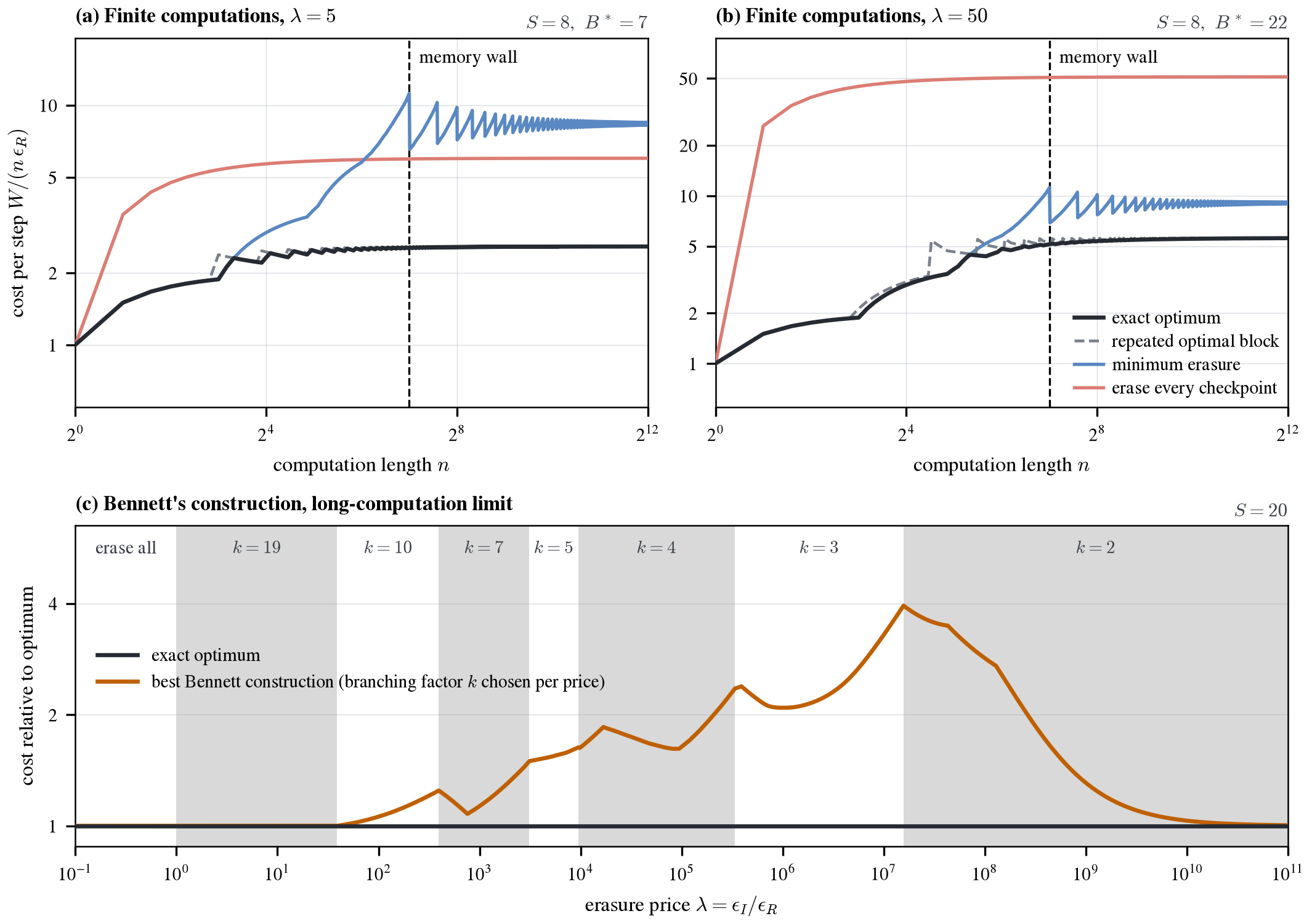}
    \caption{\label{fig:strategy_advantage}
Optimal schedule compared with simpler strategies.
(a),(b) Cost per step $C/n$ at $S=8$ for $\lambda=5$ and $50$ using the exact
optimum of Eq.~(\ref{eq:priced_dp}), the repeated optimal block of
Theorem~\ref{thm:strict_exact} ($B_{\rm str}^*=7$ and $22$),
the best minimum-erasure schedule (the $\lambda \rightarrow \infty$ optimum), and erasing every checkpoint (the two-pebble glider). The dashed line marks
the memory wall $n=2^{S-1}$.
(c) The bulk cost of Bennett's construction~\cite{bennett_timespace_1989} relative to the exact bulk optimum at $S=20$. We repeat Bennett's recursive construction in blocks and erase the boundary checkpoint between consecutive blocks. At each erasure price we choose the number of segments per level $k$ and the recursion depth $h$ that give the lowest cost per node. Shaded bands mark the selected $k$.}
\end{figure*}

\section{The bulk optimum}
While the recurrence in Eq.~(\ref{eq:priced_dp}) gives the exact
finite-length optimum, long computations have a simpler structure. Away from the endpoints, a periodic schedule attains the optimal cost per step, with each period a strictly reversible block followed by a single erasure.

Suppose the only stored checkpoint is a boundary node $x$. A strict
period of length $B$ holds $x$ fixed, uses the remaining $S-1$ pebbles to
compute node $x+B$ while reversibly cleaning every intermediate checkpoint, and then erases $x$. Node $x+B$ becomes the boundary for the next period
(Fig.~\ref{fig:bulk_schedule}). The
reversible part of the period is a fully reversible schedule of length $B$ with $S-1$
pebbles, so it costs $F(B,S-1)$ and is feasible only for
\[
B\le B_{\rm max}\equiv 2^{S-2},
\]
the clean-finish limit of $S-1$ pebbles. The erasure costs $\lambda$. One
period therefore advances the computation by $B$ nodes at a cost per node of
\begin{equation}
c_{\rm str}(B,S,\lambda)
=
\frac{F(B,S-1)+\lambda}{B}.
\label{eq:block_density}
\end{equation}
The block length trades off two effects: while increasing the block length spreads the single
erasure cost over more nodes, the reversible cost
per node $F(B,S-1)/B$ grows because cleaning a longer block with fixed memory requires more recomputation. The best block length and the resulting
density are
\begin{equation}
B_{\rm str}^*(S,\lambda)
\in
\arg\min_{1\le B\le B_{\rm max}} c_{\rm str}(B,S,\lambda),
\label{eq:b_str_star}
\end{equation}
\begin{equation}
\rho_{\rm str}(S,\lambda)
\equiv
c_{\rm str}(B_{\rm str}^*,S,\lambda).
\label{eq:rho_str}
\end{equation}

Which block length wins depends on the erasure price. Comparing two lengths
$B_2>B_1$ in Eq.~(\ref{eq:block_density}), the longer block is preferred when
\begin{equation}
\lambda\ge
\frac{B_1F(B_2,S-1)-B_2F(B_1,S-1)}{B_2-B_1}.
\label{eq:strict_pairwise_crossover}
\end{equation}
For \(\lambda\le1\), \(B=1\) is optimal and \(\rho_{\rm str}=1+\lambda\), so every
old checkpoint is erased immediately. Increasing \(\lambda\) selects
progressively longer strict blocks until \(B_{\rm str}^*\) reaches \(B_{\rm max}=2^{S-2}\). At $B_{\rm max}$, repeating the block reproduces the springboard schedule of Ref.~\cite{li_reversible_1998}, except that each period cleans up with Knill's time-optimal $F(2^{S-2},S-1)$ reversible moves rather than the $3^{S-2}$ of Bennett's recursive schedule.

\begin{lemma}[Strict-partition domination]
\label{lem:strict_partition}
Consider a priced schedule constructed recursively by
Eq.~(\ref{eq:priced_dp}) that advances \(B\) nodes with space \(k\), uses \(R\)
reversible moves, and contains \(I\) internal erasures.
There is an ordered composition separated by erasures
\[
B=b_1+\cdots+b_{I+1}
\,.
\]
Every part obeys \(b_j\le2^{k-1}\). The reversible work obeys
\begin{equation}
R\ge\sum_{j=1}^{I+1}F(b_j,k).
\label{eq:strict_partition_bound}
\end{equation}
\end{lemma}

The proof (Appendix~\ref{app:strict_partition_proof}) inducts over the tree of separators generated by Eq.~(\ref{eq:priced_dp}).

\begin{theorem}[Exact repeated-strict-block bulk optimum]
\label{thm:strict_exact}
For every \(S\ge2\) and \(\lambda\ge0\),
\begin{equation}
C_\lambda(n,S)
=\rho_{\rm str}(S,\lambda)n+O_{S,\lambda}(1).
\label{eq:finite_density_rate}
\end{equation}
Consequently \(C_\lambda(n,S)/n\to\rho_{\rm str}(S,\lambda)\) as
\(n\to\infty\). Repeating a minimizing strict block \(B_{\rm str}^*\le2^{S-2}\), followed by one local erasure, attains this exact bulk density.
\end{theorem}

\begin{proof}
Let \(k=S-1\). For any \(B\), take an optimal schedule \(C_\lambda(B,k)=R+\lambda I\) from
Theorem~\ref{thm:priced_separator_recurrence}. By Lemma
\ref{lem:strict_partition} it splits into \(I+1\) parts, each a feasible strict block, with
\(R\ge\sum_jF(b_j,k)\). Including one boundary erasure,
\begin{align}
C_\lambda(B,k)+\lambda
&=R+\lambda(I+1)\nonumber\\
&\ge\sum_{j=1}^{I+1}\bigl[F(b_j,k)+\lambda\bigr] \ge\rho_{\rm str}\sum_{j=1}^{I+1}b_j
=\rho_{\rm str}B.
\label{eq:mixed_block_domination}
\end{align}

For the upper bound, let \(B^*=B_{\rm str}^*\) and
\(n=qB^*+p\), with \(1\le p\le B^*\). Computing the remainder $p$
and then \(q\) strict periods gives
\[
C_\lambda(n,S)
\le q\rho_{\rm str}B^*+\max_{1\le p\le B^*}C_\lambda(p,S)
\le\rho_{\rm str}n+O_{S,\lambda}(1).
\]

For the matching lower bound, we set \(K=\rho_{\rm str}B_{\rm max}+\lambda\). We prove
\(C_\lambda(L,S)\ge\rho_{\rm str}L-K\) for every \(L\ge1\) by induction on \(L\).
The case \(L=1\) follows from
\(B_{\rm max}\ge1\) and nonnegative costs. For \(L\ge2\), let \(m\) minimize
Eq.~(\ref{eq:priced_dp}) and set \(B=L-m\). If \(m>B_{\rm max}\), then
\(F(m,S-1)=\infty\), so the separator is erased. The induction hypothesis and
Eq.~(\ref{eq:mixed_block_domination}) give
\[
C_\lambda(L,S)
\ge\rho_{\rm str}m-K+\rho_{\rm str}B
=\rho_{\rm str}L-K.
\]
If \(m\le B_{\rm max}\), Eq.~(\ref{eq:mixed_block_domination}) gives
\(C_\lambda(B,S-1)\ge\rho_{\rm str}B-\lambda\). Dropping the other nonnegative terms
in the recurrence and using \(B=L-m\ge L-B_{\rm max}\) again gives
\[
C_\lambda(L,S)
\ge\rho_{\rm str}B-\lambda
\ge\rho_{\rm str}L-K.
\]
The upper and lower bounds prove Eq.~(\ref{eq:finite_density_rate}).
\end{proof}

Because the endpoint correction is bounded, the cost per step approaches
$\rho_{\rm str}(S,\lambda)$ as $O(1/n)$, and Fig.~\ref{fig:strategy_advantage}(a),(b) shows the repeated block approaching the exact optimum within a few periods. For comparison, the figure
also shows the minimum-erasure schedule, which is the $\lambda\to\infty$
optimum, and the schedule that erases every checkpoint. At $\lambda=5$ the
former can cost more than the latter, so insisting on reversibility can cost
more than abandoning it.

Figure~\ref{fig:strategy_advantage}(c) compares the bulk optimum with
periods built from Bennett's construction~\cite{bennett_timespace_1989}, the classic reversible time--space tradeoff, in its general form with $k$ segments per recursion level. With $h$ levels, it advances $k^h$ nodes using $(2k-1)^h$ reversible moves and $h(k-1)+1$ pebbles. We choose $k$ and $h$ to fit within $S-1$ pebbles and minimize the bulk density at each price. A single level is a plain compute--uncompute of $k\leq S-1$ nodes, which matches the optimum whenever $B_{\rm str}^*\leq S-1$. At intermediate prices, Bennett's construction
costs up to several times the optimum. Once $\lambda$ grows large, both schedules use blocks near $B_{\rm max}$ and the erasure term dominates, so they agree again.

%%%%%%%%%%%%%%
\section{From operation prices to physical work}
\label{sec:physical_work}

Modern stochastic thermodynamics provides a broad framework for analyzing the
entropic costs of computational processes
\cite{wolpert_stochastic_2019}. Considering externally clocked thermally coupled devices whose clock and control protocols are fixed across a computation, we can map protocol timesteps $\tau$ to our relative prices $\lambda$ to find the corresponding physical work. All candidate schedules
are therefore compared at the same primitive duration $\tau$, although their
total completion times may differ because they contain different numbers of
operations.

For fixed endpoints and a time-rescaled protocol in the linear-response
regime, the leading mean excess work has the form $A/\tau$
\cite{sivak_thermodynamic_2012}. Let $\alpha\geq0$ be the finite-time
coefficient per erased bit and $\beta>0$ the coefficient of one reversible
checkpoint move. Both coefficients may themselves depend on temperature
\cite{sivak_thermodynamic_2012}, though we suppress that dependence in the notation below, because the schedule optimization depends on the hardware only through
$\lambda$. As stated in Sec.~\ref{sec:priced_game}, we assume that a discarded
checkpoint is uniformly distributed over its $2^b$ values and that resets are
local, so they cannot exploit correlations with retained states. The
quasistatic checkpoint reset cost is then the total Landauer cost
\[
E_{\rm L}(T)\equiv b k_B T\ln2 ,
\]
and we take the same reset price at every checkpoint location. Algorithm- or
data-specific resets can do better \cite{frank_foundations_2017}, but this
uniform cost keeps the schedule algorithm-independent. On the linear-response validity domain $\mathcal D=[\tau_{\min},\infty)$ we take
\begin{align}
\epsilon_I(T,\tau)
&=E_{\rm L}(T)+\frac{b\alpha}{\tau}, &\epsilon_R(\tau)
&=\frac{\beta}{\tau},
\label{eq:reversible_energy}
\end{align}
where $\beta$ may depend on both the checkpoint width $b$ and the gates
being implemented. In this model the operating point of Sec.~I is $H=(T,\tau)$, with the
coefficients $\alpha$ and $\beta$ fixed by the device.

Finite-time erasure models provide realizations of the $1/\tau$ term
\cite{zulkowski_optimal_2014,proesmans_finite-time_2020,
proesmans_optimal_2020}, and experiments on colloidal bits measure an erasure
cost approaching the Landauer value with the same $1/\tau$ correction
\cite{berut_experimental_2012,jun_high-precision_2014}. Outside $\mathcal D$, the cost law depends on the device dynamics and allowed controls
\cite{kolchinsky_work_2021,zhong_limited-control_2022,
dago_adiabatic_2023}.

\begin{figure*}[t]
\centering
\includegraphics[width=0.92\textwidth]{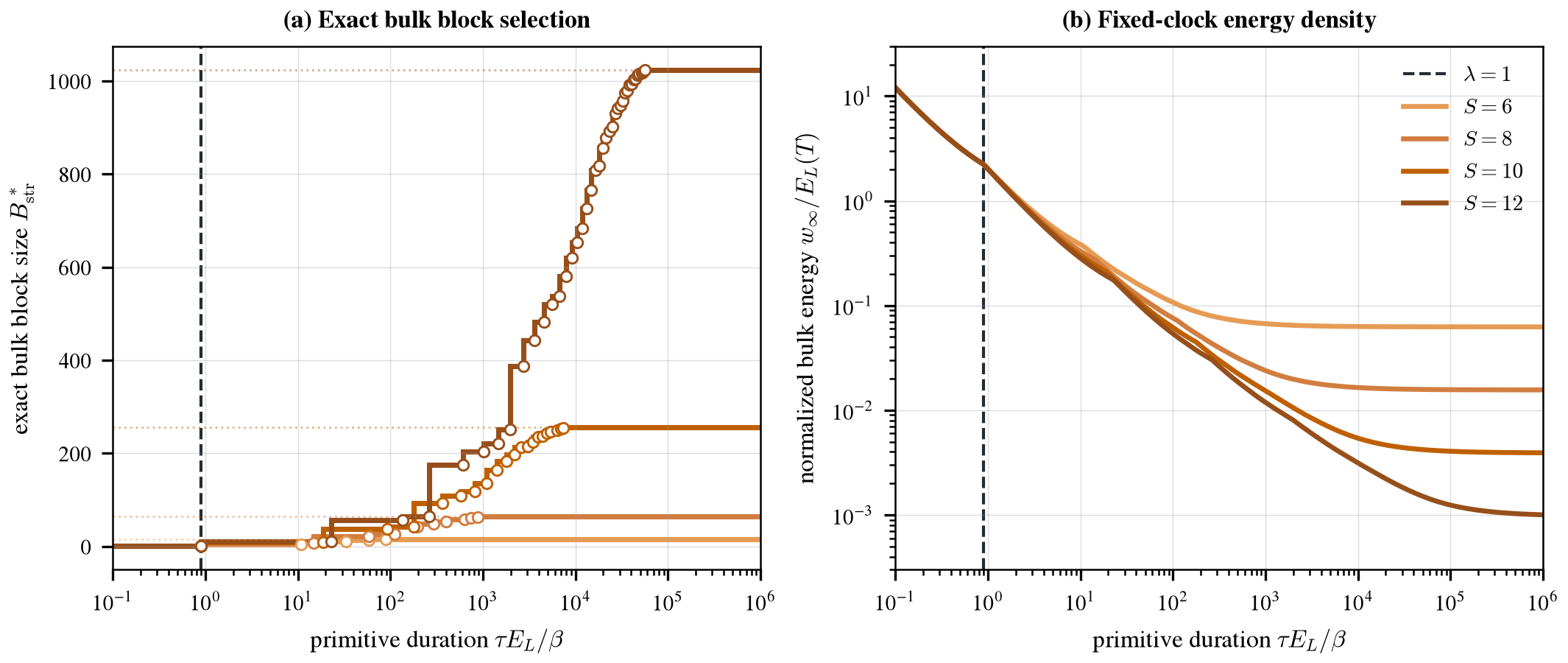}
\caption{\label{fig:bulk_energy}
Exact fixed-period bulk optimum for $S=6,8,10,12$ under
Eq.~(\ref{eq:reversible_energy}), with $b\alpha/\beta=0.1$ and the illustrative
validity domain $\tau E_{\rm L}(T)/\beta\geq0.1$.
(a) Optimal strict-block length $B_{\rm str}^*$. Dotted lines mark
$B_{\rm max}=2^{S-2}$. At an exact crossover several block lengths tie.
The curve breaks ties by minimizing the operation density
$[F(B,S-1)+1]/B$, and open markers show the shortest tied block. At
$\lambda=1$ the curve displays $B=1$.
(b) Corresponding normalized bulk work. The dashed line marks $\lambda=1$, at
$\tau E_{\rm L}/\beta=0.9$; to its left every optimal non-output disposal is an
erasure and additional checkpoint memory does not lower the work, and
reversible removals may tie on the line itself. This is an illustrative
$b\alpha<\beta$ slice of the general hardware-dependent phase diagram.}
\end{figure*}

The induced relative price is
\begin{equation}
\lambda(T,\tau)
=
\frac{\epsilon_I(T,\tau)}{\epsilon_R(\tau)}
=
\frac{\tau E_{\rm L}(T)+b\alpha}{\beta}.
\label{eq:lambda}
\end{equation}
Thus, at fixed temperature, increasing $\tau$ makes erasure more expensive
relative to reversible cleanup. An abstract crossover $\lambda_\times$ at which the optimal block length changes, given by Eq.~(\ref{eq:strict_pairwise_crossover}) for the two competing lengths, therefore occurs at the primitive duration $\tau_\times=(\beta\lambda_\times-b\alpha)/E_{\rm L}(T)$, where $B_{\rm str}^*$ jumps (Fig.~\ref{fig:bulk_energy}(a)). We restore units to find the finite-length minimum energy is
\[
W_{\min}(n,S,T,\tau)
=
\epsilon_R(\tau)C_{\lambda(T,\tau)}(n,S).
\]
Theorem~\ref{thm:strict_exact} then gives the exact bulk work per simulated
step as
\begin{equation}
\begin{aligned}
w_\infty(S,T,\tau)
&=
\min_{1\leq B\leq B_{\rm max}}
\frac{F(B,S-1)\epsilon_R(\tau)+\epsilon_I(T,\tau)}{B}
\\
&=
\min_{1\leq B\leq B_{\rm max}}
\left\{
\frac{E_{\rm L}(T)}{B}
+
\frac{\beta F(B,S-1)+b\alpha}{B\tau}
\right\}.
\end{aligned}
\label{eq:w_bulk_exact}
\end{equation} 

Within this model, $\lambda\geq b\alpha/\beta$ at every primitive duration,
with the bound approached at short durations or low temperatures. If
$b\alpha<\beta$, erasing every checkpoint is optimal whenever
$\tau E_{\rm L}(T)\leq\beta-b\alpha$, where $B_{\rm str}^*=1$ and
$w_\infty=\epsilon_R+\epsilon_I$ is independent of $S$, because extra
checkpoint memory does not help the two-pebble glider. If $b\alpha\geq\beta$,
the model never makes erasure cheaper than a reversible move.
Figure~\ref{fig:bulk_energy} uses $b\alpha/\beta=0.1$, so the boundary is at
$\tau E_{\rm L}/\beta=0.9$.

In the opposite, slow-driving limit, reversible excess work vanishes, but old
boundary checkpoints must still be erased at least once every $B_{\rm max}$
steps. Since $F(B,S-1)\geq B$, Eq.~(\ref{eq:w_bulk_exact}) is bounded below by
$\epsilon_R+\epsilon_I/B_{\rm max}$ and above by its value at
$B=B_{\rm max}$. Both bounds converge as $\tau\to\infty$ to the floor
\begin{equation}
w_{\rm floor}(S,T)=\frac{E_{\rm L}(T)}{B_{\rm max}},
\label{eq:model_checkpoint_floor}
\end{equation}
which sets the plateaus in Fig.~\ref{fig:bulk_energy}(b). Without a time constraint (see Appendix~\ref{app:fixed_deadline}), the outer minimization in Eq.~(\ref{eq:outer_optimization}) therefore favors the slowest possible clock at fixed temperature.

%%%%%%%%%%%%%%%

\section{Discussion}

We have determined the exact tradeoff between reversible checkpoint
moves and local erasures for a line computation with bounded memory and fixed operation prices. The separator recurrence gives the finite-length optimal schedule, while optimal schedules in the bulk of long computations are attained by repeating one strictly reversible block followed by one local erasure. 

At a fixed hardware operating point, the platform supplies
$\epsilon_R$, $\epsilon_I$, and a checkpoint-memory budget.
Changing the hardware operating point can reduce the cost of reversible operations
toward the adiabatic slow-driving limit, but at a fixed operating point, reversible cleanup need not minimize energy. Elective erasure can be energy optimal, since logically reversible operations will dissipate energy when performed in finite time. The scheduling results apply to any device model that supplies the assumed uniform operation prices.

Wolpert et al.\ call for understanding how energy cost trades off against time, memory, and other traditional computational resources, and attribute the gap between actual and minimal dissipation to physical constraints on real computers \cite{wolpert2024stochastic}. Demaine et al.\ similarly identify lower bounds on
time--space--energy tradeoffs as a major open direction for semi-reversible
computing \cite{demaine_energy-efficient_2016}. For line computations in the
pebble model, Theorems~\ref{thm:priced_separator_recurrence}
and~\ref{thm:strict_exact} give such a tradeoff exactly. Real computers face
many physical constraints, including clocks, memory, and limited connectivity.
Two of them, a fixed clock period and bounded checkpoint memory, already force a
nonzero dissipation per step here, even though logically reversible operation
is available. The result is algorithm-independent because we assume uniform
checkpoint widths and operation prices, uniformly distributed checkpoint
contents, and local resets.

A more detailed thermodynamic analysis requires a specified algorithm, input
distribution, and physical implementation
\cite{yadav_circuits_2025,yadav_entropy_2026}, which would also fix the
generalized Landauer cost of each reset. Allowing
the bit width and gate depth to vary along the computation, or computing on a
specific DAG, adds scheduling complexity. We also neglect routing costs,
external clocking costs, and bit storage costs. 

To maximize computational throughput per watt, spatial parallelization can be paired with slower computation up to the limits set by the serial portions of the dependency structure, as quantified by Amdahl's law \cite{konopik_fundamental_2023}. Once parallelization is exhausted, the remaining serial parts can be represented as line computations using complete-state checkpoints, to which the scheduling result here applies. Appendix~\ref{app:fixed_deadline} introduces a wall-clock deadline, which sets each schedule's primitive duration $\tau$, so its operation costs depend on its total number of moves. Jointly optimizing parallelization, clock period, and checkpoint scheduling remains an open direction.

\begin{acknowledgments}
M.C.T.\ was supported by a Department of Defense National Defense Science and Engineering Graduate (NDSEG) Fellowship. This material is based upon work supported by the Air Force Office of Scientific Research under award number FA9550-23-F-0014 in the amount of \$136,200.
D.S.\ is supported by a Schmidt Sciences Polymath Award.
\end{acknowledgments}

\appendix
 
\section{Separator normal form}
\label{app:normal_form}
 
The three-phase construction in the main text gives an upper bound on the
optimal cost. It computes a checkpoint at $m$, advances from $m$ to $n$,
and then disposes of $m$. This appendix proves the matching lower bound.
Even a schedule that interleaves work on both sides of $m$ must pay for the
same three subproblems, for a suitable choice of $m$. The proof adapts
Knill's proof of Eq.~(\ref{eq:knill_dp})
\cite[Theorem~2.1]{knill_analysis_1995} to allow for erasures. We rearrange erasures and reversible moves without increasing cost
so that an upper pebble remains present during the phase-3 lower-block cleanup. Cleanup containing erasures is treated separately because it cannot always be reversed.
 
The key question is how much memory the cleanup has. If the lower block
$\{1,\ldots,m\}$ is cleaned reversibly while an upper-block pebble is
present, cleanup has at most $S-1$ pebbles and costs at least $F(m,S-1)$.
We must therefore rule out any advantage from starting this cleanup while
the upper block is still empty, when all $S$ pebbles would be available.
Lemmas~\ref{lem:lower_block_bypass}
and~\ref{lem:clean_separator_normal_form} show that an optimal schedule can
avoid this situation. Lemma~\ref{lem:separator_normal_form} then bounds
separately the cost of building the separator, computing above it, and
cleaning below it. These are three disjoint sets of moves that make up the three phases. They need not be
three consecutive stretches of time in the original schedule.
 
Throughout, $Q_t$ is the set of active pebbles after move $t$, excluding
the permanent input at node $0$, with $Q_0=\emptyset$. We call a schedule
optimal if it has minimum cost and, among schedules of that cost, the
fewest moves. Counting moves rules out unnecessary detours even when
$\lambda=0$.
 
\begin{lemma}[Lower-block bypass]
\label{lem:lower_block_bypass}
Let $m\ge2$, $S\ge2$, and $\lambda\ge0$. Suppose
$P\subseteq\{1,\ldots,m\}$ satisfies
\[
 m\in P,\qquad P\ne\{m\},\qquad |P|\le S-1.
\]
Among legal priced schedules on this lower block from $\emptyset$ to $P$,
with space bound $S$ and $m$ protected from erasure, there is an optimal
one that never visits $\{m\}$.
\end{lemma}
 
\begin{proof}
The hypotheses imply $2\le |P|\le S-1$, so $S\ge3$. Choose an optimal
schedule $\Gamma$ with as few visits to $\{m\}$ as possible, and suppose
it visits $\{m\}$. Since $P\ne\{m\}$, there is a next move. Node $m$ is
protected from erasure, its predecessor is absent, and no nodes above $m$
are available. The only legal next move is therefore to add node $1$.
 
The move entering $\{m\}$ must remove some $q<m$. It cannot add $m$,
since adding $m$ requires $m-1$, which would then still be present. Thus
the two moves are
\[
 \{m,q\}\longrightarrow\{m\}\longrightarrow\{m,1\}.
\]
If $q=1$, they return to the same configuration and can be deleted, which
contradicts optimality.
 
If $q\ne1$, removing $q$ must be an erasure, because $q-1$ is absent.
Replace the two moves by
\[
 \{m,q\}
 \xrightarrow{\text{add }1}\{m,q,1\}
 \xrightarrow{\text{erase }q}\{m,1\}.
\]
The intermediate configuration has three pebbles, which $S\ge3$ allows.
The replacement has the same cost and move count but one fewer visit to
$\{m\}$, again a contradiction.
\end{proof}
 
\begin{lemma}[Clean-handoff normal form]
\label{lem:clean_separator_normal_form}
For $n\ge2$, $S\ge2$, and $\lambda\ge0$, there is an optimal winning
schedule in which, whenever $Q_t=\{j\}$ with $j<n$, the next move places a
pebble at $j+1$.
\end{lemma}

\begin{proof}
Call a time $t$ with $Q_t=\{j\}$, $j<n$, a \emph{bad handoff at index
$j$} if the next move is on $\{1,\ldots,j\}$. The only possible move above
$j$ from this configuration is the placement of $j+1$, so it is sufficient to
exclude bad handoffs.

Suppose, for contradiction, that every optimal winning schedule has a bad
handoff. Choose one whose largest bad-handoff index $m$ is as small as
possible. Let $t$ be its last bad handoff at index $m$, and let $v$ be the
first move above $m$ after $t$. Such a move must occur to reach $n$.

From time $t$ until just before move $v$, the upper block
$I^+=\{m+1,\ldots,n\}$ is empty, since $Q_t=\{m\}$ and $v$ is the first
upper move after $t$. Move $v$ therefore adds $m+1$. Write
$I^-=\{1,\ldots,m\}$ and $P=Q_{v-1}$. Since the upper block is empty and
$v$ is legal,
\[
 P\subseteq I^-,\qquad m\in P,\qquad |P|\le S-1.
\]
If $P=\{m\}$, the moves between $t$ and $v$ return the schedule to
$\{m\}$. There is at least one such move because $t$ is a bad handoff, so
deleting them contradicts optimality. Hence $P\ne\{m\}$. Then $|P|\ge2$,
which gives $m\ge2$ and $S\ge3$.

Keep only the lower-block moves in the prefix before $v$, preserving their
order. This gives a legal schedule from $\emptyset$ to $P$, because lower
moves do not depend on upper pebbles and removing upper moves can only
reduce occupancy. To make $m$ a protected output of this shortened
schedule, delete each placement of $m$ except the last, together with the
removal of $m$ that follows it. These deletions are legal because $m$ is a
predecessor only of $m+1$, which does not appear in the shortened
schedule. The result has no greater cost or move count and never erases
$m$.

This shortened schedule satisfies the hypotheses of
Lemma~\ref{lem:lower_block_bypass}. Replace it by an optimal one that
reaches $P$ without visiting $\{m\}$, and append the original suffix
beginning with move $v$. The result is a legal winning schedule that is at
least as good as the original, so it is also optimal.

The new prefix contains no visit to $\{m\}$ and no node above $m$, so any
bad handoff in it has index below $m$. The unchanged suffix contains no bad
handoff at $m$, because $t$ was the last one, and none above $m$, because
$m$ was the largest index. The new schedule is optimal, so by assumption it
has a bad handoff, and every bad handoff it has lies below $m$. This
contradicts the choice of $m$.
\end{proof}
 
\begin{lemma}[Separator decomposition]
\label{lem:separator_normal_form}
For \(n\ge 2\), \(S\ge 2\), and \(\lambda\ge 0\),
\begin{multline*}
C_\lambda(n,S) \ge \min_{1\le m<n} \big[ C_\lambda(m,S) + C_\lambda(n-m,S-1) \\
+ \min\{F(m,S-1),\lambda\} \big].
\end{multline*}
\end{lemma}
 
\begin{proof}
By Lemma~\ref{lem:clean_separator_normal_form}, there is an optimal winning
schedule $\Sigma$ in which, whenever $Q_t=\{j\}$ with $j<n$, the next move
places a pebble at $j+1$. Let $t_f$ be its final time, so $Q_0=\emptyset$
and $Q_{t_f}=\{n\}$.
An earlier visit to $\{n\}$ or a return to $\emptyset$ would let the
schedule be shortened, so $Q_t$ is nonempty and different from $\{n\}$
for $0<t<t_f$.
 
\emph{Choosing the separator.}
Track the leftmost occupied node and take its largest value before
completion,
\[
 \ell(t)=\min Q_t,\qquad m=\max_{0<t<t_f}\ell(t).
\]
Since $Q_t\ne\{n\}$, $\ell(t)<n$, so $1\le m<n$. Since $\ell(t)\le m$, the lower block $I^-=\{1,\ldots,m\}$ contains at least one pebble at every intermediate time. Let $I^+=\{m+1,\ldots,n\}$, and let $t_m$ be the last time before $t_f$ at which $\ell(t_m)=m$, so that
\[
 Q_{t_m}\cap I^-=\{m\}.
\]
 
The upper block is also occupied at $t_m$. Otherwise $Q_{t_m}=\{m\}$, and
Lemma~\ref{lem:clean_separator_normal_form} makes the next move add $m+1$.
That would give $Q_{t_m+1}=\{m,m+1\}$ and $\ell(t_m+1)=m$ at a later time
before $t_f$, which contradicts the choice of $t_m$. The upper block also
cannot become empty again before $t_f$. If it did, the upper-block moves
from its first occupation through that return would take it from empty
back to empty. Lower moves never depend on upper pebbles, so deleting
those upper moves would give a legal winning schedule with no greater cost
and fewer moves. Hence $I^+$ is occupied at every time from $t_m$ to
$t_f$.
 
We partition the moves into three sets. These are the lower-block moves up
to $t_m$, all upper-block moves, and the lower-block moves after $t_m$.
Each move belongs to exactly one set, and we bound their respective costs.
 
\emph{1. Building the lower checkpoint.}
Keep only the lower-block moves up to time $t_m$. They form a legal
lower-block schedule from $\emptyset$ to $\{m\}$ using at most $S$
pebbles. As in the proof of Lemma~\ref{lem:clean_separator_normal_form},
delete each placement of $m$ except the last, together with the removal of
$m$ that follows it. This protects $m$ from erasure without increasing
cost, so these moves cost at least $C_\lambda(m,S)$.
 
\emph{2. Computing the upper block.}
Keep all moves on $I^+$ in their original order and treat $m$ as their
permanent input. Every reversible move at $m+1$ originally had $m$ present, and an erasure needs no predecessor. All other predecessor conditions lie within $I^+$. This gives a legal priced schedule from the empty upper block to $\{n\}$. Because the lower block is occupied at every intermediate time, the upper block subproblem uses at most $S-1$ pebbles. Relabeling $m+1,\ldots,n$ as $1,\ldots,n-m$ therefore shows that these
moves cost at least $C_\lambda(n-m,S-1)$.
 
\emph{3. Cleaning the lower block.}
The remaining lower-block moves form a sequence $U$. It starts from
$Q_{t_m}\cap I^-=\{m\}$ and ends with the lower block empty, since
$Q_{t_f}=\{n\}$. We consider two cases. 

(i) If $U$ contains an erasure, it costs at least $\lambda$, the price of
erasing $m$ directly, because every move has nonnegative cost. No mixture of further erasures and reversible moves can bring the cleanup below this. 

(ii) If $U$ contains no erasure, every move of $U$ is reversible. Each is made after $t_m$, while $I^+$ is occupied, so the lower block holds at most
$S-1$ pebbles throughout $U$. Reversing the order of the moves and swapping placements with removals gives a legal construction of $\{m\}$ from $\emptyset$, because a reversible move needs the same predecessor in both directions. An erasure may not have a legal reverse, because its predecessor can be absent. Hence a reversible $U$ costs at least
$F(m,S-1)$. 

With the two cases at hand, we have shown that $U$ costs at least $\min\{F(m,S-1),\lambda\}$.

Adding the three bounds yields
\begin{multline*}
 \operatorname{cost}(\Sigma)\ge
 C_\lambda(m,S)+C_\lambda(n-m,S-1)\\
 {}+\min\{F(m,S-1),\lambda\}.
\end{multline*}
This holds for the separator chosen from $\Sigma$, so it also holds with
the right-hand side minimized over all $1\le m<n$.
\end{proof}
 
Lemma~\ref{lem:separator_normal_form} shows that interleaving moves cannot reduce the cost below Eq.~(\ref{eq:priced_dp}). Together with the three-phase construction in the main text, this proves Theorem~\ref{thm:priced_separator_recurrence}.

\section{Strict-partition proof}
\label{app:strict_partition_proof}
We use two standard monotonicities of the reversible cost
\cite[Observation 2.2(ii)--(iii)]{knill_analysis_1995}:
\begin{equation}
F(a,k+1)\leq F(a,k),\qquad
F(a,k)\leq F(m,k) \qquad (a\leq m).
\label{eq:F_monotonicities}
\end{equation}
The merge step below also uses Knill's reversible separator
recurrence \cite[Theorem 2.1]{knill_analysis_1995}.

\begin{proof}[{ Proof of
Lemma~\ref{lem:strict_partition}}]
 Induct on the separator tree. The one-node schedule is the
base case. Suppose the root separator splits \(B=m+d\). Let the left (phase 1) and right (phase 2)
subproblems have data \((R_1,I_1)\) at space \(k\) and \((R_2,I_2)\) at space
\(k-1\), and apply the induction hypothesis to both.

If the separator is erased in phase 3, take the left parts followed by the right parts, with the erased separator \(m\) as the boundary between them. Promoting the right parts from space \(k-1\) to \(k\) cannot
increase their reversible cost by Eq.~(\ref{eq:F_monotonicities}). Given $I = I_1+I_2+1$, the
composition then has
\((I_1+1)+(I_2+1)=I+1\) parts and strict cost at most
\(R_1+R_2=R\).

 If the separator is cleaned reversibly, its cleanup costs
\(F(m,k-1)\) and \(I=I_1+I_2\). Let \(a\) be the last left part and \(d_1\)
the first right part. Feasibility gives \(m\le2^{k-2}\) and \(a\le m\).
The space-\((k-1)\) induction hypothesis also gives
\(d_1\le2^{k-2}\), and thus \(a+d_1\le2^{k-1}\).
The Knill recurrence and Eq.~(\ref{eq:F_monotonicities}) give

\begin{equation}
 \begin{aligned}
F(a+d_1,k)
&\le F(a,k)+F(d_1,k-1)+F(a,k-1)\\
&\le F(a,k)+F(d_1,k-1)+F(m,k-1).
\end{aligned}
\label{eq:strict_partition_merge}
\end{equation}
So the actual cleanup pays for merging the two boundary
parts into \(a+d_1\). All remaining right parts can be promoted to space
\(k\). The resulting composition has \(I+1\) parts, total length \(B\), and
strict cost at most \(R_1+R_2+F(m,k-1)=R\).
\end{proof}

\section{Scheduling under a fixed completion time}
\label{app:fixed_deadline}

The main text compares schedules at a fixed primitive period
\(\tau\). Suppose instead that a serialized computation must finish by an
externally imposed time \(\mathcal T\), with every primitive in the selected
schedule uniformly retimed. If a schedule \(\sigma\) uses \(I\) erasures and \(R\) reversible
moves, write \(N=I+R\). The cost of any fixed schedule decreases with $\tau$,
so an optimal schedule uses the entire available time, $\tau_\sigma=\mathcal T / N$. We restrict to schedules with
$N\le\mathcal T/\tau_{\min}$, so that $\tau_\sigma\in\mathcal D$.

Substitution into Eq.~(\ref{eq:reversible_energy}) gives
\[
W_{\mathcal T}(\sigma)
=E_{\rm L}(T)I
+\frac{N}{\mathcal T}\left(b\alpha I+ \beta R\right).
\]
Recalling that $\mathcal A_{n,S}$ is the set of attainable
count pairs $(I(\sigma),R(\sigma))$ for $\sigma\in\mathcal F_{n,S}$, the exact
serialized fixed-completion-time problem is
\begin{equation}
W_{\mathcal T}^*(n,S,T)
=
\min_{\substack{(I,R)\in\mathcal A_{n,S}}}
\left[
E_{\rm L}(T)I
+\frac{I+R}{\mathcal T}\left(b\alpha I +\beta R\right)
\right].
\label{eq:fixed_deadline_exact}
\end{equation}

Since the objective increases with both $I$ and $R$, the optimum is a
Pareto-minimal pair in $\mathcal A_{n,S}$. Each schedule runs at its own
period $\mathcal T/N$, however, so the optimum need not minimize the
fixed-price cost at any single $\lambda$.

\bibliography{references_final}

@article{lent_bennett_2006,
	title = {Bennett clocking of quantum-dot cellular automata and the limits to binary logic scaling},
	volume = {17},
	issn = {0957-4484},
	url = {https://doi.org/10.1088/0957-4484/17/16/040},
	doi = {10.1088/0957-4484/17/16/040},
	number = {16},
	urldate = {2026-04-17},
	journal = {Nanotechnology},
	author = {Lent, Craig S and Liu, Mo and Lu, Yuhui},
	month = aug,
	year = {2006},
	pages = {4240}
}

@article{proesmans_optimal_2020,
	title = {Optimal finite-time bit erasure under full control},
	volume = {102},
	url = {https://link.aps.org/doi/10.1103/PhysRevE.102.032105},
	doi = {10.1103/PhysRevE.102.032105},
	number = {3},
	urldate = {2026-08-14},
	journal = {Physical Review E},
	publisher = {American Physical Society},
	author = {Proesmans, Karel and Ehrich, Jannik and Bechhoefer, John},
	month = sep,
	year = {2020},
	pages = {032105}
}

@article{kolchinsky_work_2021,
	title = {Work, {Entropy} {Production}, and {Thermodynamics} of {Information} under {Protocol} {Constraints}},
	volume = {11},
	url = {https://link.aps.org/doi/10.1103/PhysRevX.11.041024},
	doi = {10.1103/PhysRevX.11.041024},
	number = {4},
	urldate = {2026-08-13},
	journal = {Physical Review X},
	publisher = {American Physical Society},
	author = {Kolchinsky, Artemy and Wolpert, David H.},
	month = nov,
	year = {2021},
	pages = {041024}
}

@article{zhong_limited-control_2022,
	title = {Limited-control optimal protocols arbitrarily far from equilibrium},
	volume = {106},
	url = {https://link.aps.org/doi/10.1103/PhysRevE.106.044135},
	doi = {10.1103/PhysRevE.106.044135},
	number = {4},
	urldate = {2026-08-13},
	journal = {Physical Review E},
	publisher = {American Physical Society},
	author = {Zhong, Adrianne and DeWeese, Michael R.},
	month = oct,
	year = {2022},
	pages = {044135}
}

@article{li_reversible_1998,
	title = {Reversible {Simulation} of {Irreversible} {Computation} by {Pebble} {Games}},
	volume = {120},
	issn = {01672789},
	url = {http://arxiv.org/abs/quant-ph/9703009},
	doi = {10.1016/S0167-2789(98)00052-9},
	number = {1-2},
	urldate = {2026-06-09},
	journal = {Physica D: Nonlinear Phenomena},
	author = {Li, Ming and Tromp, John and Vit{\'a}nyi, Paul},
	month = sep,
	year = {1998},
	note = {arXiv:quant-ph/9703009},
	pages = {168--176}
}

@article{kralovic_time_2004,
	title = {Time and space complexity of reversible pebbling},
	volume = {38},
	issn = {0988-3754, 1290-385X},
	url = {https://www.cambridge.org/core/journals/rairo-theoretical-informatics-and-applications/article/abs/time-and-space-complexity-of-reversible-pebbling/895043387916FF807C5CF086195B861C},
	doi = {10.1051/ita:2004008},
	number = {2},
	urldate = {2026-03-11},
	journal = {RAIRO - Theoretical Informatics and Applications},
	author = {Kr{\'a}lovi{\v{c}}, Richard},
	month = apr,
	year = {2004},
	pages = {137--161}
}

@article{wolpert_thermodynamics_2020,
	title = {Thermodynamics of computing with circuits},
	volume = {22},
	issn = {1367-2630},
	url = {https://doi.org/10.1088/1367-2630/ab82b8},
	doi = {10.1088/1367-2630/ab82b8},
	number = {6},
	urldate = {2026-06-01},
	journal = {New Journal of Physics},
	publisher = {IOP Publishing},
	author = {Wolpert, David H and Kolchinsky, Artemy},
	month = jun,
	year = {2020},
	pages = {063047}
}

@article{sivak_thermodynamic_2012,
	title = {Thermodynamic {Metrics} and {Optimal} {Paths}},
	volume = {108},
	copyright = {http://link.aps.org/licenses/aps-default-license},
	issn = {0031-9007, 1079-7114},
	url = {https://link.aps.org/doi/10.1103/PhysRevLett.108.190602},
	doi = {10.1103/PhysRevLett.108.190602},
	number = {19},
	urldate = {2026-03-13},
	journal = {Physical Review Letters},
	author = {Sivak, David A. and Crooks, Gavin E.},
	month = may,
	year = {2012},
	pages = {190602}
}

@article{bennett_thermodynamics_1982,
	title = {The thermodynamics of computation---a review},
	volume = {21},
	copyright = {http://www.springer.com/tdm},
	issn = {0020-7748, 1572-9575},
	url = {https://doi.org/10.1007/BF02084158},
	doi = {10.1007/BF02084158},
	number = {12},
	urldate = {2026-03-15},
	journal = {International Journal of Theoretical Physics},
	author = {Bennett, Charles H.},
	month = dec,
	year = {1982},
	pages = {905--940}
}

@article{wolpert_stochastic_2019,
	title = {The stochastic thermodynamics of computation},
	volume = {52},
	issn = {1751-8121},
	url = {https://doi.org/10.1088/1751-8121/ab0850},
	doi = {10.1088/1751-8121/ab0850},
	number = {19},
	urldate = {2026-06-01},
	journal = {Journal of Physics A: Mathematical and Theoretical},
	publisher = {IOP Publishing},
	author = {Wolpert, David H},
	month = apr,
	year = {2019},
	pages = {193001}
}

@article{wolpert2024stochastic,
	title = {Is stochastic thermodynamics the key to understanding the energy costs of computation?},
	volume = {121},
	number = {45},
	doi = {10.1073/pnas.2321112121},
	url = {https://doi.org/10.1073/pnas.2321112121},
	journal = {Proceedings of the National Academy of Sciences},
	author = {Wolpert, David H. and Korbel, Jan and Lynn, Christopher W. and Tasnim, Farita and Grochow, Joshua A. and Karde{\c{s}}, G{\"u}lce and Aimone, James B. and Balasubramanian, Vijay and De Giuli, Eric and Doty, David and Freitas, Nahuel and Marsili, Matteo and Ouldridge, Thomas E. and Richa, Andr{\'e}a W. and Riechers, Paul and Rold{\'a}n, {\'E}dgar and Rubenstein, Brenda and Toroczkai, Zoltan and Paradiso, Joseph},
	year = {2024},
	pages = {e2321112121}
}

@article{faist_minimal_2015,
	title = {The minimal work cost of information processing},
	volume = {6},
	copyright = {2015 The Author(s)},
	issn = {2041-1723},
	url = {https://www.nature.com/articles/ncomms8669},
	doi = {10.1038/ncomms8669},
	number = {1},
	urldate = {2026-06-01},
	journal = {Nature Communications},
	publisher = {Nature Publishing Group},
	author = {Faist, Philippe and Dupuis, Frédéric and Oppenheim, Jonathan and Renner, Renato},
	month = jul,
	year = {2015},
	pages = {7669}
}

@article{li_reversibility_1996,
	author = {Li, Ming and Vit{\'a}nyi, Paul},
	title = {Reversibility and adiabatic computation: trading time and space for energy},
	volume = {452},
	issn = {1364-5021, 1471-2946},
	shorttitle = {Reversibility and adiabatic computation},
	url = {https://royalsocietypublishing.org/doi/10.1098/rspa.1996.0039},
	doi = {10.1098/rspa.1996.0039},
	number = {1947},
	urldate = {2026-03-11},
	journal = {Proceedings of the Royal Society of London. Series A: Mathematical, Physical and Engineering Sciences},
	month = apr,
	year = {1996},
	pages = {769--789}
}

@article{konopik_fundamental_2023,
	title = {Fundamental energy cost of finite-time parallelizable computing},
	volume = {14},
	copyright = {2023 The Author(s)},
	issn = {2041-1723},
	url = {https://www.nature.com/articles/s41467-023-36020-2},
	doi = {10.1038/s41467-023-36020-2},
	number = {1},
	urldate = {2026-06-02},
	journal = {Nature Communications},
	publisher = {Nature Publishing Group},
	author = {Konopik, Michael and Korten, Till and Lutz, Eric and Linke, Heiner},
	month = jan,
	year = {2023},
	pages = {447}
}

@inproceedings{frank_foundations_2017,
	address = {Cham},
	title = {Foundations of {Generalized} {Reversible} {Computing}},
	isbn = {978-3-319-59936-6},
	doi = {10.1007/978-3-319-59936-6_2},
	booktitle = {Reversible {Computation}},
	publisher = {Springer International Publishing},
	author = {Frank, Michael P.},
	editor = {Phillips, Iain and Rahaman, Hafizur},
	year = {2017},
	pages = {19--34}
}

@article{proesmans_finite-time_2020,
	title = {Finite-{Time} {Landauer} {Principle}},
	volume = {125},
	issn = {0031-9007, 1079-7114},
	url = {https://link.aps.org/doi/10.1103/PhysRevLett.125.100602},
	doi = {10.1103/PhysRevLett.125.100602},
	number = {10},
	urldate = {2026-03-13},
	journal = {Physical Review Letters},
	author = {Proesmans, Karel and Ehrich, Jannik and Bechhoefer, John},
	month = sep,
	year = {2020},
	pages = {100602}
}

@article{fredkin_conservative_1982,
	title = {Conservative logic},
	volume = {21},
	issn = {1572-9575},
	url = {https://doi.org/10.1007/BF01857727},
	doi = {10.1007/BF01857727},
	number = {3},
	urldate = {2026-03-16},
	journal = {International Journal of Theoretical Physics},
	author = {Fredkin, Edward and Toffoli, Tommaso},
	month = apr,
	year = {1982},
	pages = {219--253}
}

@article{levine_note_1990,
	title = {A note on {Bennett}'s time-space tradeoff for reversible computation},
	volume = {19},
	number = {4},
	journal = {SIAM Journal on Computing},
	author = {Levine, Robert Y. and Sherman, Alan T.},
	doi = {10.1137/0219046},
	year = {1990},
	pages = {673--677}
}

@article{bennett_timespace_1989,
	title = {Time/{Space} {Trade}-{Offs} for {Reversible} {Computation}},
	volume = {18},
	issn = {0097-5397},
	url = {https://epubs.siam.org/doi/10.1137/0218053},
	doi = {10.1137/0218053},
	number = {4},
	urldate = {2026-06-02},
	journal = {SIAM Journal on Computing},
	publisher = {Society for Industrial and Applied Mathematics},
	author = {Bennett, Charles H.},
	month = aug,
	year = {1989},
	pages = {766--776}
}

@misc{herr_landauer-limited_2025,
	title = {Landauer-{Limited} {Dissipation} in {Quantum}-{Flux}-{Parametron} {Logic}},
	url = {http://arxiv.org/abs/2504.04284},
	doi = {10.48550/arXiv.2504.04284},
	urldate = {2026-06-01},
	publisher = {arXiv},
	author = {Herr, Quentin},
	month = apr,
	year = {2025},
	note = {arXiv:2504.04284 [physics.app-ph]}
}

@article{kolchinsky_dependence_2021,
	title = {Dependence of integrated, instantaneous, and fluctuating entropy production on the initial state in quantum and classical processes},
	volume = {104},
	url = {https://link.aps.org/doi/10.1103/PhysRevE.104.054107},
	doi = {10.1103/PhysRevE.104.054107},
	number = {5},
	urldate = {2026-06-01},
	journal = {Physical Review E},
	publisher = {American Physical Society},
	author = {Kolchinsky, Artemy and Wolpert, David H.},
	month = nov,
	year = {2021},
	pages = {054107}
}

@article{yadav_entropy_2026,
	title = {Entropy production bounds for systems running computer programs},
	volume = {5},
	issn = {2752-6542},
	url = {https://doi.org/10.1093/pnasnexus/pgag116},
	doi = {10.1093/pnasnexus/pgag116},
	number = {4},
	urldate = {2026-06-01},
	journal = {PNAS Nexus},
	author = {Yadav, Abhishek and Caravelli, Francesco and Wolpert, David},
	month = apr,
	year = {2026},
	pages = {pgag116}
}

@misc{yadav_circuits_2025,
	title = {Minimal thermodynamic cost of computing with circuits},
	url = {https://arxiv.org/abs/2504.04031},
	doi = {10.48550/arXiv.2504.04031},
	publisher = {arXiv},
	author = {Yadav, Abhishek and Yousef, Mahran and Wolpert, David},
	year = {2025},
	note = {arXiv:2504.04031 [cond-mat.stat-mech]}
}

@article{kamijima_finite-time_2025,
	title = {Finite-time thermodynamic bounds and trade-off relations for information processing},
	volume = {7},
	url = {https://journals.aps.org/prresearch/abstract/10.1103/PhysRevResearch.7.013329},
	doi = {10.1103/PhysRevResearch.7.013329},
	number = {1},
	journal = {Physical Review Research},
	publisher = {American Physical Society},
	author = {Kamijima, Takuya and Funo, Ken and Sagawa, Takahiro},
	month = mar,
	year = {2025},
	pages = {013329}
}

@inproceedings{demaine_energy-efficient_2016,
	address = {New York, NY, USA},
	series = {{ITCS} '16},
	title = {Energy-{Efficient} {Algorithms}},
	isbn = {978-1-4503-4057-1},
	url = {https://dl.acm.org/doi/10.1145/2840728.2840756},
	doi = {10.1145/2840728.2840756},
	urldate = {2026-06-01},
	booktitle = {Proceedings of the 2016 {ACM} {Conference} on {Innovations} in {Theoretical} {Computer} {Science}},
	publisher = {Association for Computing Machinery},
	author = {Demaine, Erik D. and Lynch, Jayson and Mirano, Geronimo J. and Tyagi, Nirvan},
	month = jan,
	year = {2016},
	pages = {321--332}
}

@article{yamae_minimum_2024,
	title = {Minimum energy dissipation required for information processing using adiabatic quantum-flux-parametron circuits},
	volume = {135},
	issn = {0021-8979},
	url = {https://doi.org/10.1063/5.0187756},
	doi = {10.1063/5.0187756},
	number = {6},
	urldate = {2026-04-17},
	journal = {Journal of Applied Physics},
	author = {Yamae, Taiki and Takeuchi, Naoki and Yoshikawa, Nobuyuki},
	month = feb,
	year = {2024},
	pages = {063902}
}

@article{bennett_notes_2003,
	title = {Notes on {Landauer}'s principle, reversible computation, and {Maxwell}'s {Demon}},
	volume = {34},
	issn = {1355-2198},
	url = {https://www.sciencedirect.com/science/article/pii/S135521980300039X},
	doi = {10.1016/S1355-2198(03)00039-X},
	number = {3},
	urldate = {2026-04-14},
	journal = {Studies in History and Philosophy of Science Part B: Studies in History and Philosophy of Modern Physics},
	author = {Bennett, Charles H.},
	month = sep,
	year = {2003},
	pages = {501--510}
}

@misc{knill_analysis_1995,
	title = {An analysis of {Bennett}'s pebble game},
	url = {http://arxiv.org/abs/math/9508218},
	doi = {10.48550/arXiv.math/9508218},
	urldate = {2026-03-13},
	publisher = {arXiv},
	author = {Knill, Emanuel},
	month = aug,
	year = {1995},
	note = {arXiv:math/9508218}
}

@article{zulkowski_optimal_2014,
	title = {Optimal finite-time erasure of a classical bit},
	volume = {89},
	url = {https://link.aps.org/doi/10.1103/PhysRevE.89.052140},
	doi = {10.1103/PhysRevE.89.052140},
	number = {5},
	urldate = {2026-03-11},
	journal = {Physical Review E},
	publisher = {American Physical Society},
	author = {Zulkowski, Patrick R. and DeWeese, Michael R.},
	month = may,
	year = {2014},
	pages = {052140}
}

@article{landauer_irreversibility_1961,
	title = {Irreversibility and {Heat} {Generation} in the {Computing} {Process}},
	volume = {5},
	issn = {0018-8646},
	url = {https://ieeexplore.ieee.org/document/5392446/},
	doi = {10.1147/rd.53.0183},
	number = {3},
	urldate = {2026-03-11},
	journal = {IBM Journal of Research and Development},
	author = {Landauer, R.},
	month = jul,
	year = {1961},
	pages = {183--191}
}

@article{bennett_logical_1973,
	title = {Logical {Reversibility} of {Computation}},
	volume = {17},
	issn = {0018-8646},
	url = {https://ieeexplore.ieee.org/document/5391327/},
	doi = {10.1147/rd.176.0525},
	number = {6},
	urldate = {2026-03-11},
	journal = {IBM Journal of Research and Development},
	author = {Bennett, C. H.},
	month = nov,
	year = {1973},
	pages = {525--532}
}

@article{dago_adiabatic_2023,
	title = {Adiabatic computing for optimal thermodynamic efficiency of information processing},
	volume = {120},
	url = {https://www.pnas.org/doi/abs/10.1073/pnas.2301742120},
	doi = {10.1073/pnas.2301742120},
	number = {39},
	urldate = {2025-06-09},
	journal = {Proceedings of the National Academy of Sciences},
	publisher = {Proceedings of the National Academy of Sciences},
	author = {Dago, Salambô and Ciliberto, Sergio and Bellon, Ludovic},
	month = sep,
	year = {2023},
	pages = {e2301742120}
}

@article{berut_experimental_2012,
	title = {Experimental verification of {Landauer}’s principle linking information and thermodynamics},
	volume = {483},
	copyright = {2012 Springer Nature Limited},
	issn = {1476-4687},
	url = {https://www.nature.com/articles/nature10872},
	doi = {10.1038/nature10872},
	number = {7388},
	urldate = {2025-06-09},
	journal = {Nature},
	publisher = {Nature Publishing Group},
	author = {Bérut, Antoine and Arakelyan, Artak and Petrosyan, Artyom and Ciliberto, Sergio and Dillenschneider, Raoul and Lutz, Eric},
	month = mar,
	year = {2012},
	pages = {187--189}
}

@article{koomey_implications_2011,
	title = {Implications of {Historical} {Trends} in the {Electrical} {Efficiency} of {Computing}},
	volume = {33},
	issn = {1058-6180},
	doi = {10.1109/MAHC.2010.28},
	number = {3},
	journal = {IEEE Annals of the History of Computing},
	author = {Koomey, Jonathan and Berard, Stephen and Sanchez, Marla and Wong, Henry},
	year = {2011},
	pages = {46--54}
}

@article{lent_device_1997,
	title = {A device architecture for computing with quantum dots},
	volume = {85},
	issn = {1558-2256},
	url = {https://ieeexplore.ieee.org/document/573740},
	doi = {10.1109/5.573740},
	number = {4},
	urldate = {2026-10-12},
	journal = {Proceedings of the IEEE},
	author = {Lent, C.S. and Tougaw, P.D.},
	month = apr,
	year = {1997},
	pages = {541--557}
}

@article{kornerup_spooky_2025,
	title = {Tight bounds on the spooky pebble game: recycling qubits with measurements},
	volume = {9},
	doi = {10.22331/q-2025-02-18-1636},
	journal = {Quantum},
	author = {Kornerup, Niels and Sadun, Jonathan and Soloveichik, David},
	year = {2025},
	pages = {1636}
}

@article{takeuchi_reversibility_2017,
	title = {Reversibility and energy dissipation in adiabatic superconductor logic},
	volume = {7},
	doi = {10.1038/s41598-017-00089-9},
	journal = {Scientific Reports},
	author = {Takeuchi, Naoki and Yamanashi, Yuki and Yoshikawa, Nobuyuki},
	year = {2017},
	pages = {75}
}

@inproceedings{frank_reversible_2020,
	title = {Reversible computing with fast, fully static, fully adiabatic {CMOS}},
	booktitle = {2020 {International} {Conference} on {Rebooting} {Computing} ({ICRC})},
	publisher = {IEEE},
	author = {Frank, Michael P. and Brocato, Robert W. and Tierney, Brian D. and Missert, Nancy A. and Hsia, Alexander H.},
	year = {2020},
	note = {arXiv:2009.00448}
}

@article{jun_high-precision_2014,
	title = {High-precision test of {Landauer}'s principle in a feedback trap},
	volume = {113},
	number = {19},
	doi = {10.1103/PhysRevLett.113.190601},
	journal = {Physical Review Letters},
	author = {Jun, Yonggun and Gavrilov, Mom{\v{c}}ilo and Bechhoefer, John},
	year = {2014},
	pages = {190601}
}

@article{zhen_universal_2021,
	title = {Universal bound on energy cost of bit reset in finite time},
	volume = {127},
	number = {19},
	doi = {10.1103/PhysRevLett.127.190602},
	journal = {Physical Review Letters},
	author = {Zhen, Yi-Zheng and Egloff, Dario and Modi, Kavan and Dahlsten, Oscar},
	year = {2021},
	pages = {190602}
}

@article{lee_speed_2022,
	title = {Speed limit for a highly irreversible process and tight finite-time {Landauer}'s bound},
	volume = {129},
	number = {12},
	doi = {10.1103/PhysRevLett.129.120603},
	journal = {Physical Review Letters},
	author = {Lee, Jae Sung and Lee, Sangyun and Kwon, Hyukjoon and Park, Hyunggyu},
	year = {2022},
	pages = {120603}
}

\end{document}